\documentclass[%
reprint,%
 aip,pof%
]{revtex4-1}

\let\oldvec\vec 
\usepackage{amsmath}%
\usepackage{amssymb}%
\usepackage{docs}%
\usepackage{bm}%
\usepackage[colorlinks=true,linkcolor=blue]{hyperref}%
\expandafter\ifx\csname package@font\endcsname\relax\else
 \expandafter\expandafter
 \expandafter\usepackage
 \expandafter\expandafter
 \expandafter{\csname package@font\endcsname}%
\fi
\usepackage[]{graphicx} 
\usepackage{subfig}

\usepackage{color}
\usepackage{soul}
\definecolor{red}{rgb}{1,0,0}
\definecolor{green}{rgb}{0,1,0}
\definecolor{blue}{rgb}{0,0,1}
\definecolor{gray}{rgb}{0.9,0.9,0.9}

\begin{document}
\title{Dynamic stall of an airfoil under tailored three-dimensional inflow conditions}%

\author{Dominik Traphan}
\email{dominik.traphan@uni-oldenburg.de}
\affiliation{ForWind, Institute of Physics, University of Oldenburg, 26129 Oldenburg, Germany}
\author{Tom T. B. Wester}
\affiliation{ForWind, Institute of Physics, University of Oldenburg, 26129 Oldenburg, Germany}
\author{Matthew Melius}
\affiliation{Department of Mechanical and Materials Engineering, Portland State University, Portland, Oregon 97207, USA}
\author{Joachim Peinke}
\affiliation{ForWind, Institute of Physics, University of Oldenburg, 26129 Oldenburg, Germany}
\author{Gerd G\"ulker}
\affiliation{ForWind, Institute of Physics, University of Oldenburg, 26129 Oldenburg, Germany}
\author{Ra\'{u}l Bayo\'{a}n Cal}
\affiliation{Department of Mechanical and Materials Engineering, Portland State University, Portland, Oregon 97207, USA}

\date{\today}
\revised{--}%

\begin{abstract}
	Rotor blades of wind turbines in the atmospheric boundary layer regularly experience the aerodynamic phenomenon of dynamic stall consisting of a temporary overshoot of lift and detrimental fatigue loads. Particularly the formation of dynamic stall under three-dimensional inflow conditions raises open questions. Aerodynamic behavior of a DU\hbox{~}91\hbox{-}W2\hbox{-}250 wind profile undergoing light dynamic stall is thus analyzed in a wind tunnel. Effects of a gust with streamwise and spanwise periodic variation are investigated by comparing total and local lift generation with flow formation above the airfoil. The observed stall cycle is divided into five stages of which one reveals lift overshoot of up to $16~\%$. The aerodynamic response of the airfoil shows a delay of about $1/8$ period between evolution of local angle of attack and lift giving a counterclockwise dynamic polar. 
	A proper orthogonal decomposition (POD) analysis of the flow field contributes to understand aerodynamic consequences of the three-dimensional gust. Local inflow, total lift as well as certain lift events are captured by one POD eigenmode, respectively. Obtained results lead to the conclusion that the flow and particularly the stalled wake of an airfoil facing a three-dimensional gust are strongly coupled in the spanwise direction. This yields to flow stabilization, inhibition of stall, and in turn, counterclockwise dynamic polar along with augmented total lift.
\end{abstract}

\maketitle

\section{Introduction}\label{sec:intro}
The turbulent atmospheric boundary layer (ABL) consists of a broad range of flow structures. Structures related to a fast increase of wind speed are commonly referred to as gusts. Particularly rotor blades of wind turbines suffer from gusts as they induce dynamic loads. These loads are partially absorbed by the rotor blades and the tower, go through the drive train and the power electronics, and culminate in the power grid. Dynamic loads thus negatively affect the fatigue life of the whole wind turbine \cite{Spinato2009}.

An aerodynamic phenomenon often associated with dynamic loads is dynamic stall. It is characterized by a delay of stall and a corresponding lift overshoot resulting from rapid changes of angle of attack (AoA) towards values beyond the static stall angle. 
Dynamic stall manifests differently depending on details of the experimental conditions. One major distinction is the formation of deep and light dynamic stall which differs in the respective separation behavior \cite{Mulleners2012}. In the case of light dynamic stall, flow separates successively starting at the trailing edge until the separation region extends over a significant area on the airfoil. Whereas for deep dynamic stall, massive flow separation is initiated at once yielding a separation region that spans the whole airfoil \cite{McCroskey1982a}. Up to now, there is no consensus about how major lift overshoot is generated. Either vortices associated with massive flow separation lead to dynamic stall \cite{McCroskey1981,Ericsson1988}, or delayed fluid response prior to the observation of any vortical structure \cite{Albertson1987}.

At first, dynamic stall was observed on a two-dimensional G\"ottingen 459 profile \cite{Kramer1932} and became a topic of major interest in the course of progressing helicopter development \cite{Harris1968,Liiva1969}, where the retreating blade experiences dynamic stall in forward flight. 
Over the last years, different aspects of dynamic stall have been investigated in the field of wind energy \cite{Schreck2005,SimaoFerreira2009,Choudhry2016}. The AoA acting locally at a rotor blade is determined by the ratio of wind speed and rotational velocity of the blade. When yaw of the wind turbine is misaligned with the main wind direction, rotor blades undergo periodically changing AoA \cite{Schreck2001}. This also happens in conditions where a stable, vertical gradient in wind speed is present in the ABL \cite{Oke2002}. In an extreme case, the turbine operates in complex terrain and is possibly misaligned to the main inflow that even contains wakes of other wind turbines or obstacles. This situation results in locally jumping, nearly discontinuous AoA along each rotor blade. Corresponding gusts are three-dimensional and often modulate the local AoA of the rotor blade towards values that induce dynamic stall. 

The relevance of the three-dimensional, unsteady ABL in wind energy was repeatedly shown in a stochastic sense \cite{Bottcher2007,Milan2013,Melius2014}. Though the aerodynamic impact of three-dimensional inflow on dynamic stall is mentioned in several studies \cite{Carr1988,Piziali1994,Choudhry2014}, it has not been investigated quantitatively. The present study thus aims at contributing to close this knowledge gap by experimentally characterizing aerodynamics of an airfoil subjected to a three-dimensional, periodic gust with streamwise and spanwise variation in a wind tunnel. The investigated airfoil is a DU\hbox{~}91\hbox{-}W2\hbox{-}250 profile which is specifically designed for wind energy application and widely used in mid-span portions of rotor blades \cite{Timmer2003,VanRooij2003}. This profile shows trailing edge separation once the stall AoA is exceeded. The presented experimental design thus targets light dynamic stall induced by inflow fluctuations.

In order to characterize the present form of light dynamic stall, experimental investigation is conducted as follows. After describing the experimental setup in Section \ref{sec:setup}, at first, the inflow is described in Section \ref{sec:1inflow}. The flow topology forming above the airfoil is then characterized in Section \ref{sec:2topo}. Flow is visualized by time-resolved stereoscopic particle image velocimetry (TRSPIV). Stall stages are distinguished and related to three-dimensional flow features. Then, the aerodynamic performance is investigated based on both, local lift, which is derived from the surface pressure distribution around the airfoil, and total lift obtained by force measurements (Section \ref{sec:3aero}). Finally, a proper orthogonal decomposition (POD) is used to extract dominating flow features from the highly dynamic multi-scale flow and relate fundamental flow behavior to aerodynamic performance (Section \ref{sec:4pod}). Results are discussed in Section \ref{sec:5disc} and summarized in Section \ref{sec:6concl}.

\section{Experimental setup}\label{sec:setup}
\begin{figure*}[t]
	\centering
	\captionsetup[subfigure]{justification=RaggedRight,singlelinecheck=off}
	\subfloat[justification=raggedleft][Schematic representation of three sub setups, \#1--\#3, consisting of: (a) vertically mounted DU\hbox{~}91\hbox{-}W2\hbox{-}250 profile with aspect ratio $AR= s / c = 805 {\rm~mm} / 300 {\rm~mm} = 2.68$, view on suction side, (b) active grid in vertical axes excitation mode, (c) two TRSPIV cameras in stereoscopic configuration, (d) multi-component scale, (e) surface pressure taps around the airfoil, (f) fluorescent black foil for reflection mitigation, (g) two stepper motors for setting AoA and moving axes of active grid.]{
		\label{fig:0windtunnel-a}
		\includegraphics[width=0.5\linewidth]{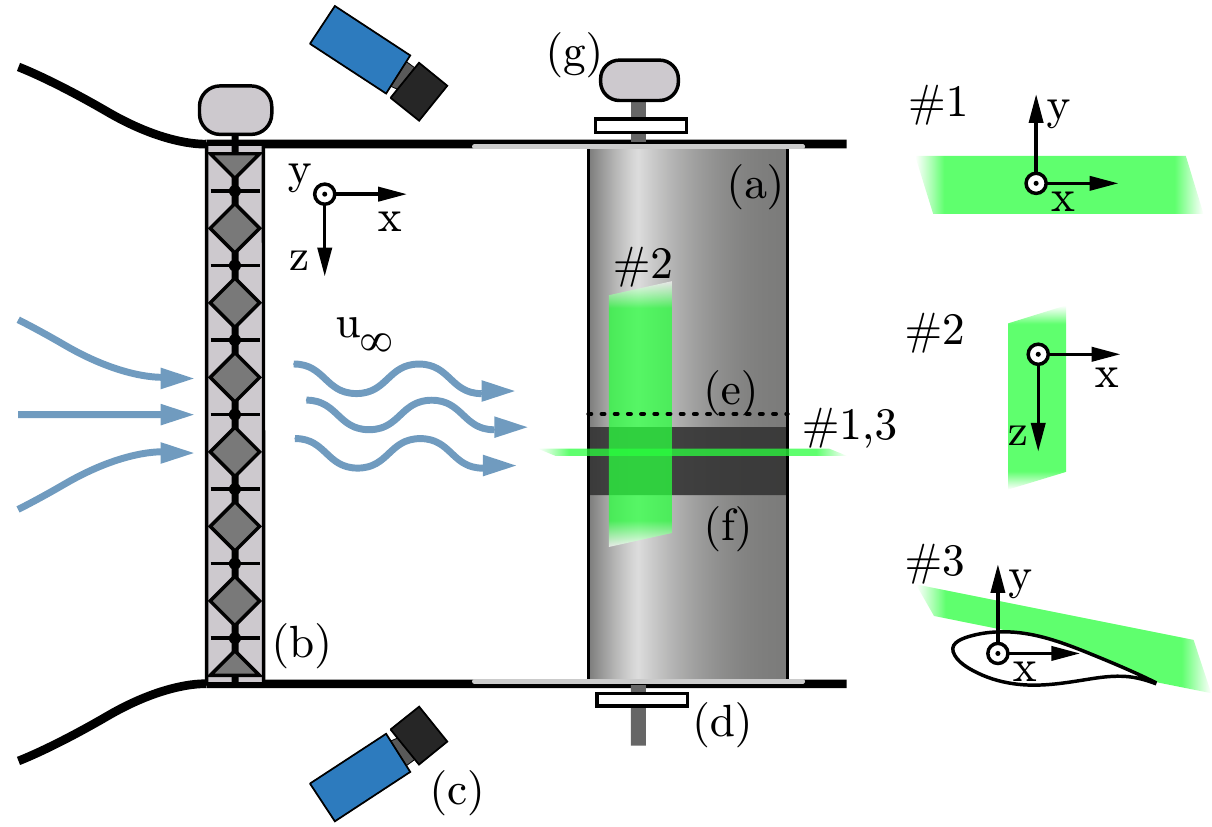}
	}
	~~~
	\subfloat[][Photograph of setup, view is directed upstream. Depiction of airfoil, active grid, upper TRSPIV camera and light-sheet configured according to configuration \#3. The airfoil with dimensions of $s \times c = 805 \times 300~{\rm mm}^2$ is located $\Delta x / c \approx 3.68$ downstream of the active grid.]{
		\label{fig:0windtunnel-b}
		\includegraphics[width=0.4\linewidth]{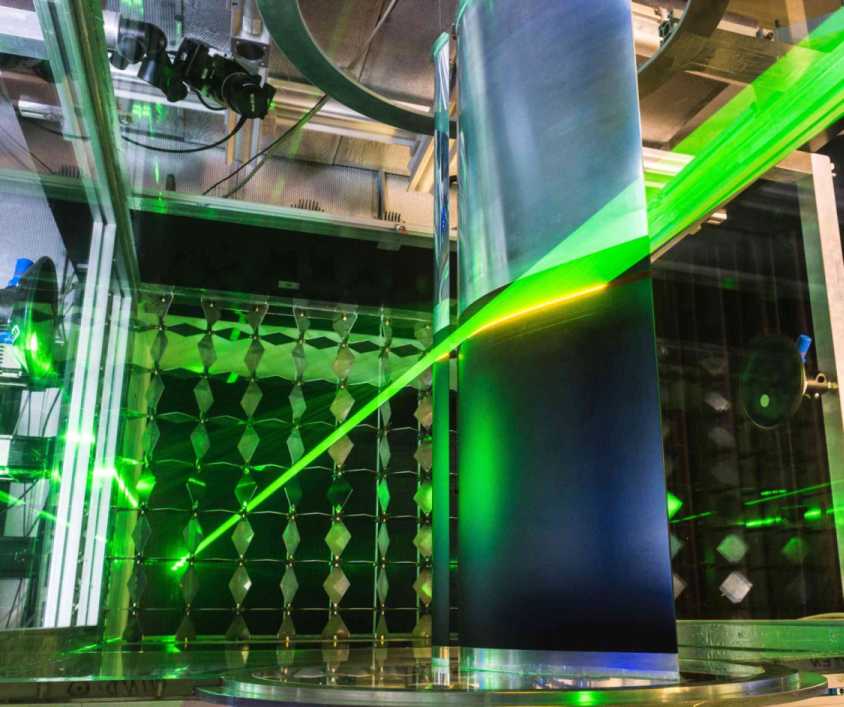}
	}
	\caption[]{Experimental setup for aerodynamic characterization of a DU\hbox{~}91\hbox{-}W2\hbox{-}250 profile in a wind tunnel.}
	\label{fig:0windtunnel}
\end{figure*}

In order to assess the impact of three-dimensional inflow on light dynamic stall, a DU\hbox{~}91\hbox{-}W2\hbox{-}250 wind profile is characterized in a closed-return wind tunnel. An active grid generates reproducible gusts tailored in time and space \cite{Weitemeyer2013}. A periodic gust with streamwise and spanwise variation is thus applied to the airfoil and investigated by means of TRSPIV, surface pressure and force measurements. This experimental setup is sketched in Figure \ref{fig:0windtunnel}\subref{fig:0windtunnel-a} as well as illustrated by a photograph showing the airfoil, parts of the TRSPIV system and the active grid in Figure \ref{fig:0windtunnel}\subref{fig:0windtunnel-b}.



The airfoil has an aspect ratio of $AR=s/c = 805 {\rm~mm} / 300 {\rm~mm} = 2.68$, where $s$ denotes the span of the airfoil and $c$ its chord length. With a mean free stream velocity of $u_\infty = 20~{\rm m/s}$, this gives a Reynolds number of $Re_{c} = 4 \times 10^5$. The test section is of $3.33~c \times 2.68~c$ in cross section and the quarter chord line of the airfoil is located $\Delta x/c = 3.68$ downstream of the wind tunnel nozzle. The coordinate origin is defined as the intercept of quarter-chord line of the airfoil and center line of the wind tunnel. The streamwise direction corresponds to $x$, wall-normal direction to $y$ and spanwise direction to $z$. At $z/c=0$, the airfoil features 30 flush pressure taps that capture the surface pressure distribution in chordwise direction. The local aerodynamic performance is determined by a spatial integration of the pressure distribution. Because of a significant tubing length to the pressure transducers, the local lift signal is obtained up to a frequency of $f=10~{\rm Hz}$ (Nyquist theorem respected). Total lift is measured by a multi-component scale up to $f=1~{\rm kHz}$ limited to eigenfrequencies of the airfoil support.

Inflow is periodically modulated in streamwise and spanwise directions by means of an active grid. The grid consists of seven horizontal and nine vertical axes individually driven by a stepper motor and equipped with nine and seven rhombic flaps, respectively. Each flap has an edgewise diameter of roughly $0.36~c$. The horizontal axes remain still and oriented with minimal inflow blockage for the purpose of this study. Vertical axes are controlled in specific sinusoidal motion that generates angle of inflow (AoI) fluctuations with a periodicity in the streamwise and spanwise directions at the location of the airfoil. The excitation frequency of the active grid $f_{\rm ex}=5~{\rm Hz}$ with the inflow velocity $u_{\infty}$ yield a reduced frequency of $f_{\rm r}=\pi f_{\rm ex} c / u_{\infty}=0.25$ \cite{Leishman2016}. Details about the functionality of the active grid are found in previous studies \cite{Knebel2011,Weitemeyer2013}.

The flow topology of generated inflow as well as the resulting flow formation on the suction side of the airfoil are visualized by means of TRSPIV. The technique is applied at two locations in three configurations, as illustrated in Figure \ref{fig:0windtunnel}\subref{fig:0windtunnel-a}. Configurations \#1 and \#3 cover a horizontal plane in $x$-$y$-directions at a $z$-height one-half flap diameter below the center line at $z/c=-0.18$. The difference between these configurations is the presence of the airfoil. 
In the vertical plane of configuration \#2, the free inflow is characterized in $x$-$z$-directions at $y/c=0$. To analyze the flow spatiotemporally, two cameras of the type Phantom Miro320S enable a measurement frequency of up to $f_s = 694~{\rm Hz}$ at a full resolution of $1920 \times 1200~{\rm px^2}$. This yields 1822 consecutive snapshots using the whole memory and a total measurement time of $t_{\rm tot}=2.6~{\rm s}$. The illumination is provided by a double cavity laser system Litron LDY303 with a maximum pulse energy of $E = 22.5~{\rm mJ/pulse}$. This allows a spatial resolution of $3\times10^{-3}~c$ at a sufficient measurement accuracy (i.e. ${\rm stereo~residue}<0.5~{\rm px}$) and a temporal resolution of 10 snapshots while flow is advecting over one chord length.

\section{Results} 
\subsection{The free stream}\label{sec:1inflow}
\begin{figure}[t]
	\centering
	\includegraphics[width=0.6\linewidth]{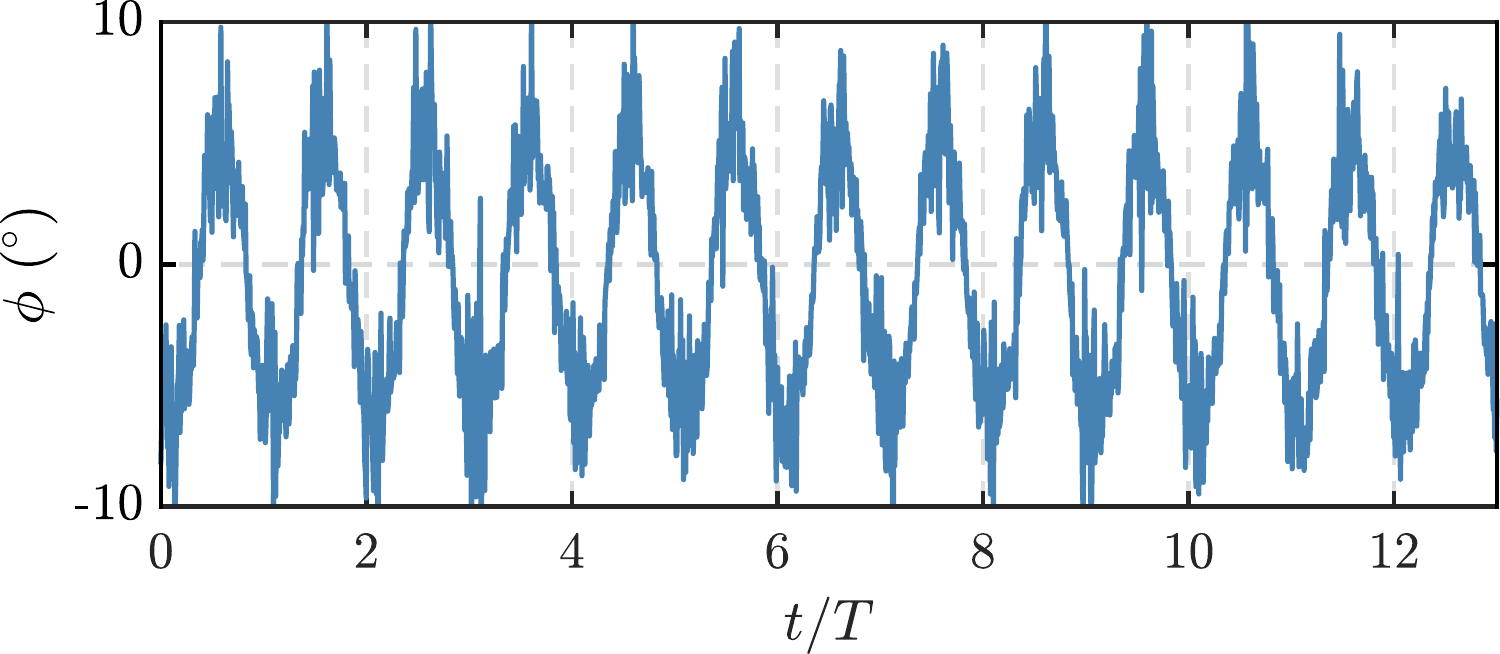}
	\caption{Evolution of AoI at intended airfoil position without airfoil present.}
	\label{fig:1InflowAlpha14}
\end{figure}

The free inflow without airfoil is first characterized. Based on TRSPIV time series of configuration \#1 ($x$-$y$-plane at $z/c=-0.18$), the free inflow reveals a turbulence intensity of $Ti = \sigma(|u|) / \langle|u|\rangle = 0.6/18.8\approx 3~\%$ in the velocity magnitude signal $|u|$. 
The AoI has an amplitude of $\phi \leq 10^{\circ}$ at the intended quarter chord point. This causes attached as well as stalled flow situation for the airfoil with a geometric mean AoA of $\alpha=10^{\circ}$. The AoI evolution $\phi(t/{T})$, where $T=200~{\rm ms}$ denotes the period length, shows the desired periodicity of $f_{\rm ex}=5~{\rm Hz}$ in Figure \ref{fig:1InflowAlpha14}. An airfoil exposed to such inflow is thus expected to encounter dynamic stall \cite{Leishman2016}.

\begin{figure}[t]
	\centering
	\includegraphics[width=0.6\linewidth]{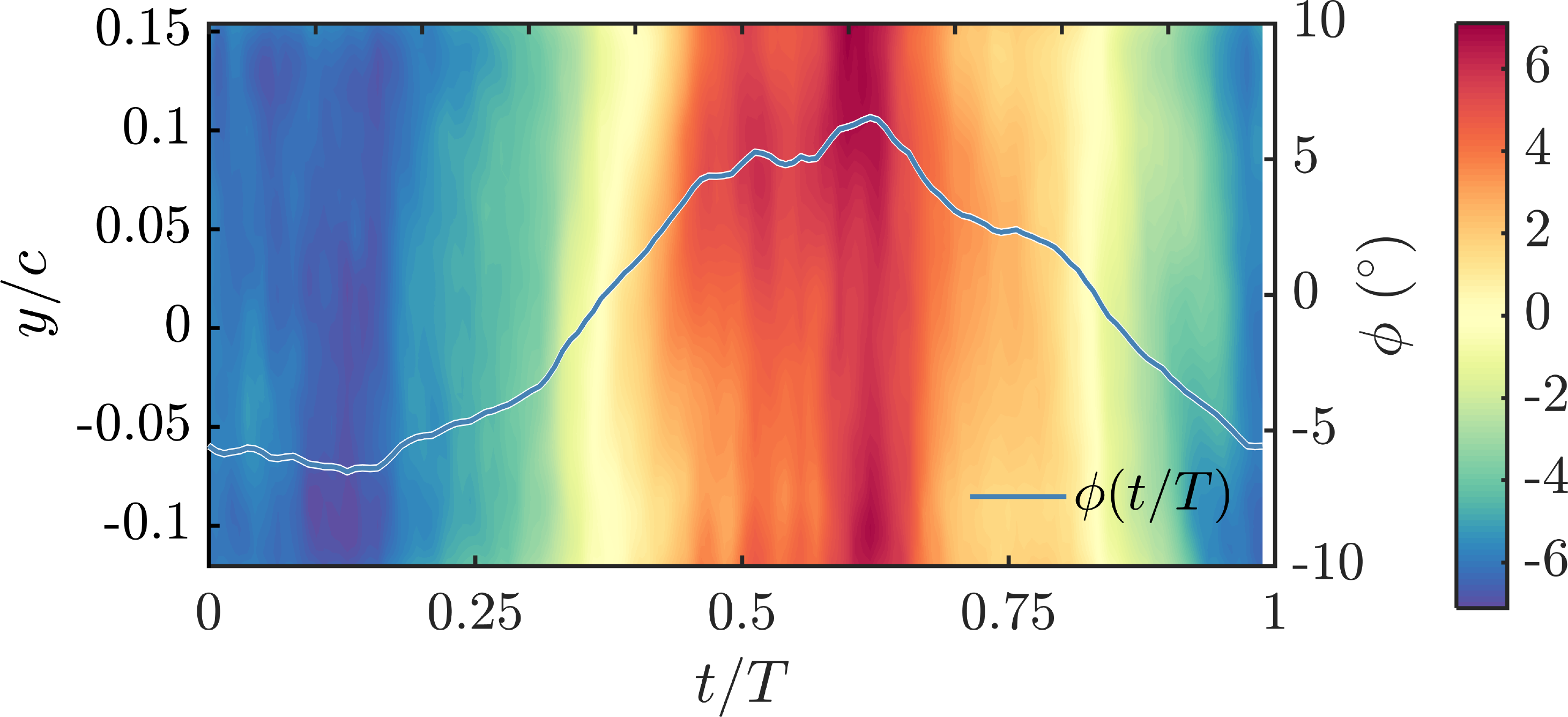}
	\caption{Evolution of AoI distributed along $y$-direction without airfoil measured in configuration \#1. The spatial average of AoI is shown as a line plot, which is used to determine effective AoA of the airfoil.}
	\label{fig:1InflowAlphaH}
\end{figure}

The AoI, averaged in streamwise direction $x$ (configuration \#1) and then phase averaged in time, is shown in Figure \ref{fig:1InflowAlphaH}. Nearly vertical ``lines'' of each color imply simultaneously changing flow properties along $y$-direction. The depicted area of $\Delta y/c=0.28$ covers the thickness of the airfoil ($\Delta y/c=0.25$) and almost one mesh size of the active grid ($\Delta y/c=0.36$). The flow pattern is used to temporally determine the local AoA for the measurements with airfoil.
This is done by averaging the AoI distribution also in the lateral $y$-direction. The evolution of this AoI is shown as a line plot in Figure \ref{fig:1InflowAlphaH}. Although the active grid operates in a perfect sinusoidal manner, the flow reveals uneven but still periodic behavior with a mean amplitude of $\langle \phi \rangle_{\rm max} \approx 6^{\circ}$ at position of the airfoil.

\begin{figure}
	\centering
	\includegraphics[width=0.6\linewidth]{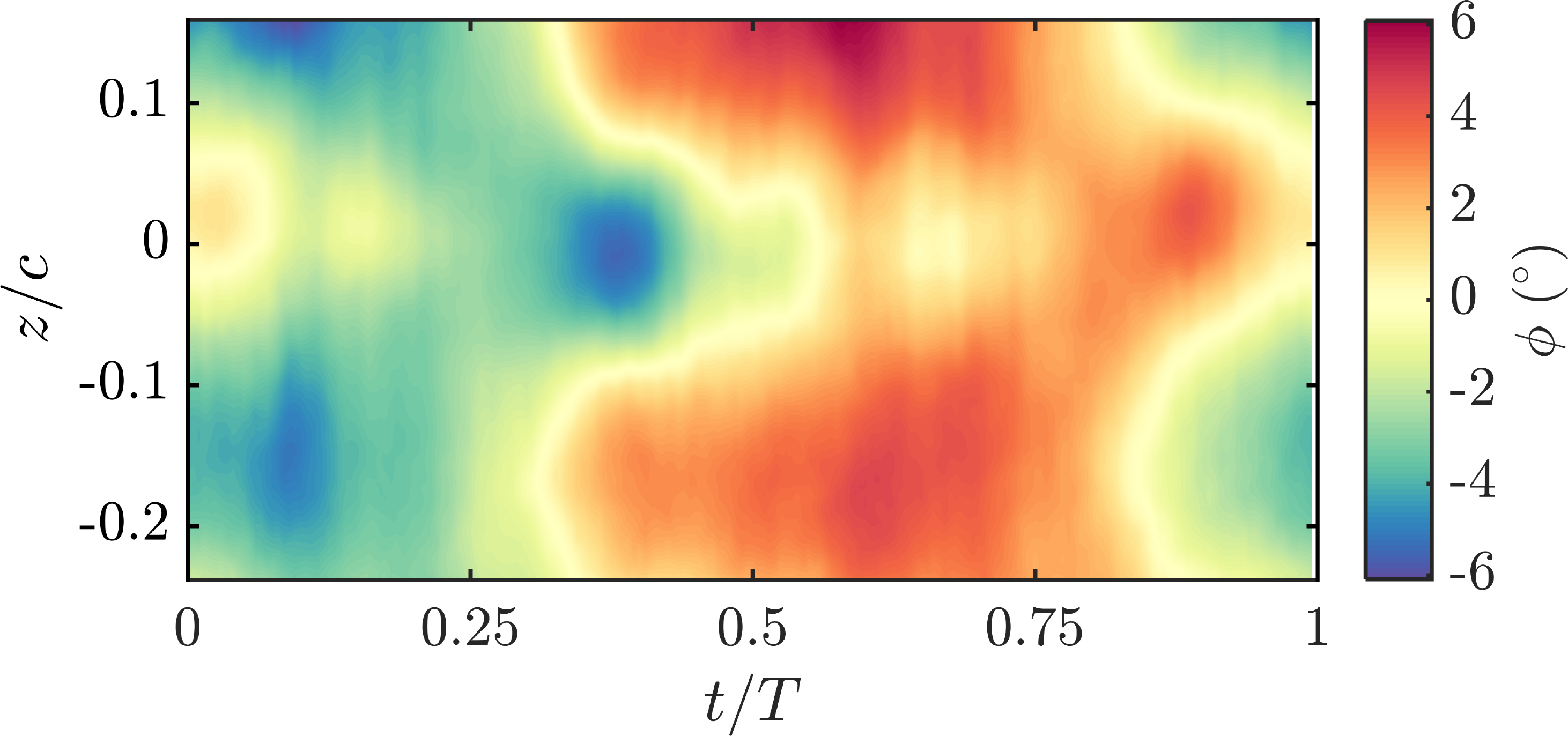}
	\caption{Evolution of AoI distributed along $z$-direction without airfoil measured in configuration \#2 (spanwise/streamwise).}
	\label{fig:1InflowAlphaV}
\end{figure}

To investigate dynamic stall in three-dimensional inflow, the inflow has to exhibit inflow variations in chordwise and in spanwise directions. The inflow characteristics in spanwise direction are thus captured by means of TRSPIV in configuration \#2. Figure \ref{fig:1InflowAlphaV} shows the evolution of AoI in $z$-direction which is the spanwise direction when the airfoil is present.

A constant AoI in $z$-direction, for example $\phi = 0^\circ$ corresponds to the color yellow in Figure \ref{fig:1InflowAlphaV}, reveals the desired periodic shape in space. The zero crossing occurs at $z/c = \pm 0.18$ at first and then moves towards $z/c = 0$ over time. An airfoil in this inflow thus encounters periodic variation of AoI along its span. Together with the inflow behavior shown in Figure \ref{fig:1InflowAlphaH}, the active grid generates spatiotemporal periodic gusts.

\subsection{Local flow topology about the airfoil}\label{sec:2topo}
\begin{figure}[t]
	\centering
	\includegraphics[width=0.6\linewidth]{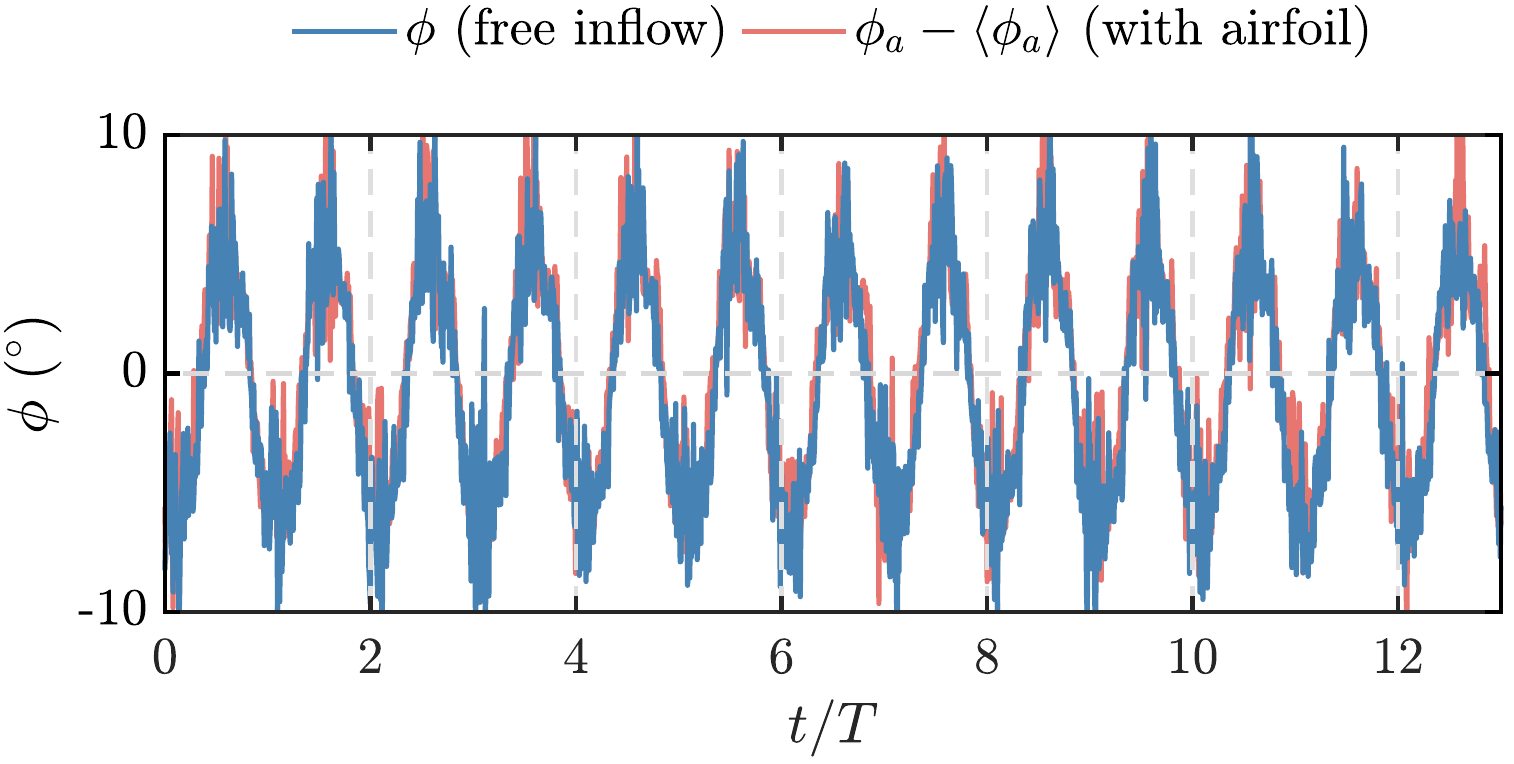}
	\caption{Evolution of AoI without airfoil $\phi$ at the intended quarter-chord point location compared to AoI with airfoil $\phi_{\rm a}$ at an upstream location above the airfoil.}
	\label{fig:2phialpha}
\end{figure}

\paragraph{Inflow conditions with airfoil:}
The evolution of AoA needs to be known to relate the flow formation to lift. The definition of AoA is however ambiguous in unsteady airfoil aerodynamics. In setups where the direction of inflow is given by the wind tunnel and the airfoil is pitched on a shaft (e.g. about quarter-chord line), the geometrical AoA with respect to the airfoil chord is biased since the leading edge of the airfoil undergoes motion that superimposes on the inflow velocity vector \cite{Choudhry2014}. In case of a fixed airfoil being exposed to fluctuating inflow, the airfoil induces an up- and downwash that can affect the temporal and/or spatial evolution of the inflow. To quantify this effect, the free inflow at the location $(x/c, y/c)=(0, 0)$ is compared with the flow that forms above the airfoil at $(x/c, y/c)=(-0.28, 0.18)$. The AoI time series of the two inflow situations, $\phi$ of the free inflow and $\phi_{\rm a}$ with airfoil, are shown in Figure \ref{fig:2phialpha} as a function of time. The AoI of the free inflow is depicted in blue. The flow angle in presence of the airfoil is plotted in red as zero-mean removing the bias of upwash. Both curves collapse within the limits of experimental noise (correlation coefficient: $\rho \approx 0.98$, signals smoothed over $c / u_\infty = 15~{\rm ms}$). Consequently, the airfoil does not significantly modify the spatiotemporal fluctuations of AoI within the three-dimensional free inflow. This finding is supported by former studies\cite{Gharali2012}, where measurements of a moving airfoil are compared with a fluctuating inflow. The respective AoA, $\alpha$, is thus defined by the geometric AoA of the airfoil, which is $10^\circ$ in this study, plus the instantaneous AoI, $\phi$, in free inflow at the quarter-chord point: $\alpha(t/ T):=\phi(t/ T)+10^{\circ}$.

\begin{figure}
	\centering
	\includegraphics[width=0.6\linewidth]{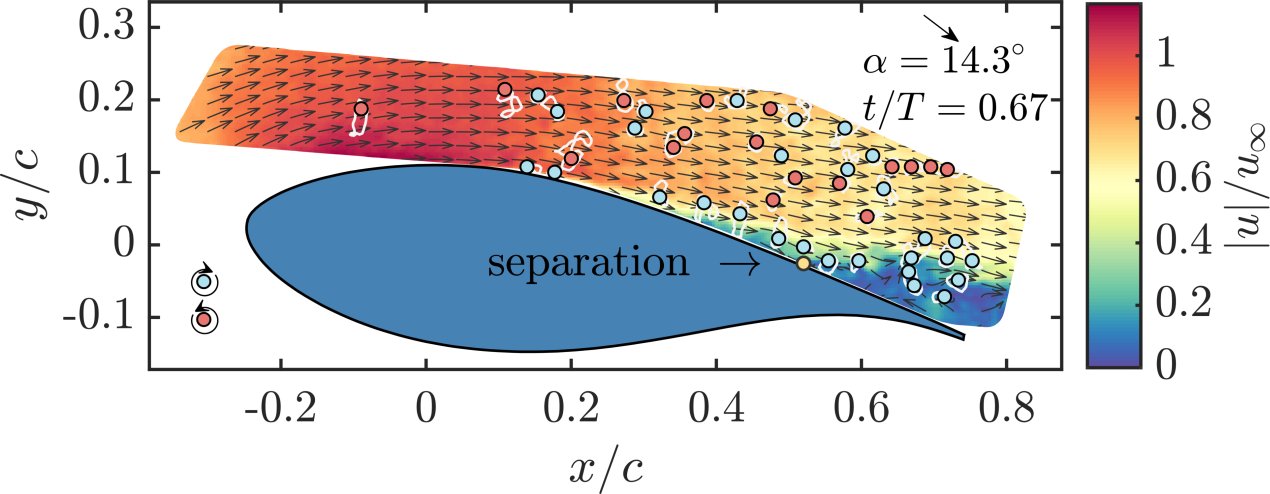}
	\caption{Snapshot of characteristic flow formation during downstroke where the flow is separated from the airfoil. The ambient flow is populated with vortices that represent instabilities. This measurement is obtained by TRSPIV configuration \#3.}
	\label{fig:2detFlow}
\end{figure}

In order to understand how the complex flow topology above the airfoil affects aerodynamic behavior, the flow field is analyzed in detail based on vortex formation and separation behavior. This procedure is illustrated in Figure \ref{fig:2detFlow} that shows the exemplary flow field above the DU\hbox{~}91\hbox{-}W2\hbox{-}250 in AoA downstroke (indicated by the arrow) at $t/T=0.67$. The magnitude of flow velocity is color-coded and vectors represent the flow direction. The airfoil is oriented with a geometric AoA of $\alpha=10^\circ$ with respect to the mean flow in $x$-direction. The separation point is defined by the location of flow reversal ($u_{\rm x}<0~{\rm m/s}$) and marked with a yellow spot. Vortices present in the flow are determined following former studies \cite{Melius2016}. Vortex centers are depicted according to $\Gamma$ criterion \cite{Graftieaux2001,Melius2018} with corresponding vortex borders in white according to zero-crossing of $\lambda_2$ criterion \cite{Jeong1995}. The color of identified vortices indicates the sense of rotation, i.e. blue and red for clockwise and counterclockwise rotating vortices, respectively.

As generally known from an airfoil undergoing light stall, Figure \ref{fig:2detFlow} shows how the flow is accelerated on the suction side until the thickest point of the airfoil, decelerates afterwards along with increasing receptivity for perturbations that provoke flow separation followed by a recirculation region. The flow above the airfoil is thus populated by clockwise and counterclockwise rotating vortices for $x/c>0.1$ reflecting the baseline turbulence intensity of $Ti \approx 3~\%$ of the free inflow.

\paragraph{Characteristic stages of 3D light dynamic stall:}
In case of deep dynamic stall, the stall cycle consists of different stages that culminate in the formation of large-scale flow structures referred to as primary and secondary dynamic stall vortices. These vortices form, advect over the airfoil and temporally generate a high lift overshoot. Deep stall occurs after separation of these massive flow structures \cite{Mulleners2012}. On the other hand, light dynamic stall implies dynamic lift overshoot coming along with successive trailing edge separation. As a consequence of early reversing inflow, which happens for high reduced frequencies and/or small fluctuation amplitudes, the dynamic stall vortex is not able to form and the premature recirculation zone is forced to reattach to the airfoil. Significant lift overshoot is observed at AoA exceeding the static stall angle because flow inertia delays separation. To differentiate these two standard dynamic stall situations from dynamic stall induced by periodic, three-dimensional inflow, the velocity fields of the present experiment are divided into characteristic topological stages.

\begin{figure}
	\centering
	\captionsetup[subfigure]{position=top,justification=RaggedRight,singlelinecheck=off}
	\subfloat[justification=raggedleft][Attached flow during upstroke.]{
		\label{fig:2topoAtt-a}
		\includegraphics[height=3.35cm]{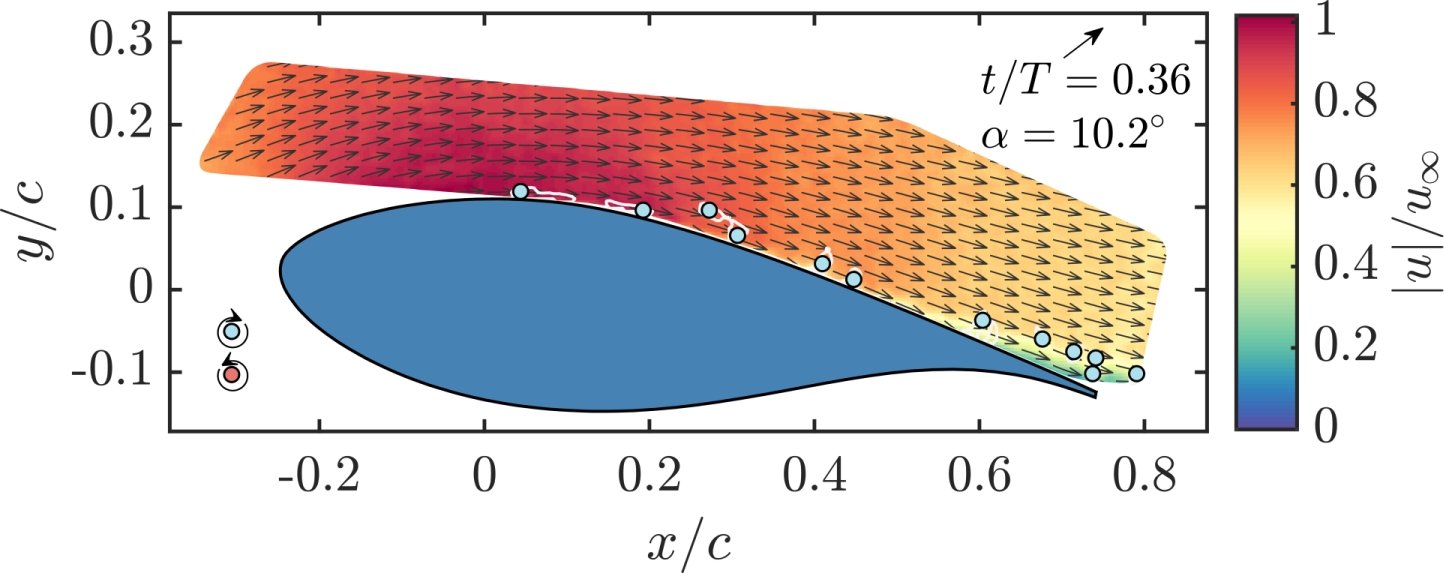}
	}
	
	\subfloat[][Onset of stall during upstroke.]{
		\label{fig:2topoOnset-b}
		\includegraphics[height=3.35cm]{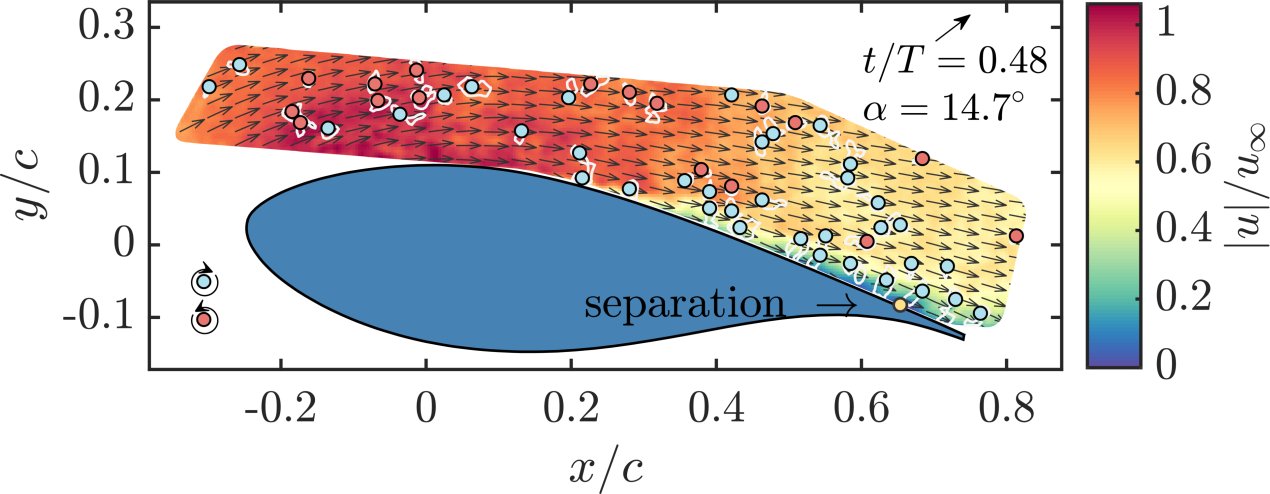}
	}
	
	\subfloat[][Maximum separation after $\alpha_{\rm max}$.]{
		\label{fig:2topoMax-c}
		\includegraphics[height=3.35cm]{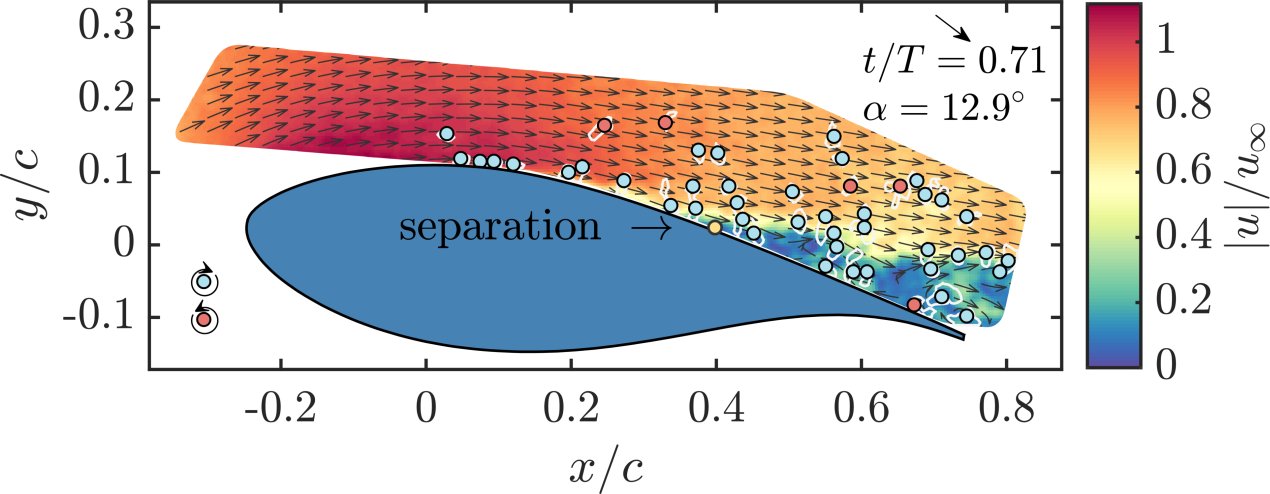}
	}
	
	\subfloat[][Forced shedding of recirculation region during downstroke.]{
		\label{fig:2topoForced-d}
		\includegraphics[height=3.35cm]{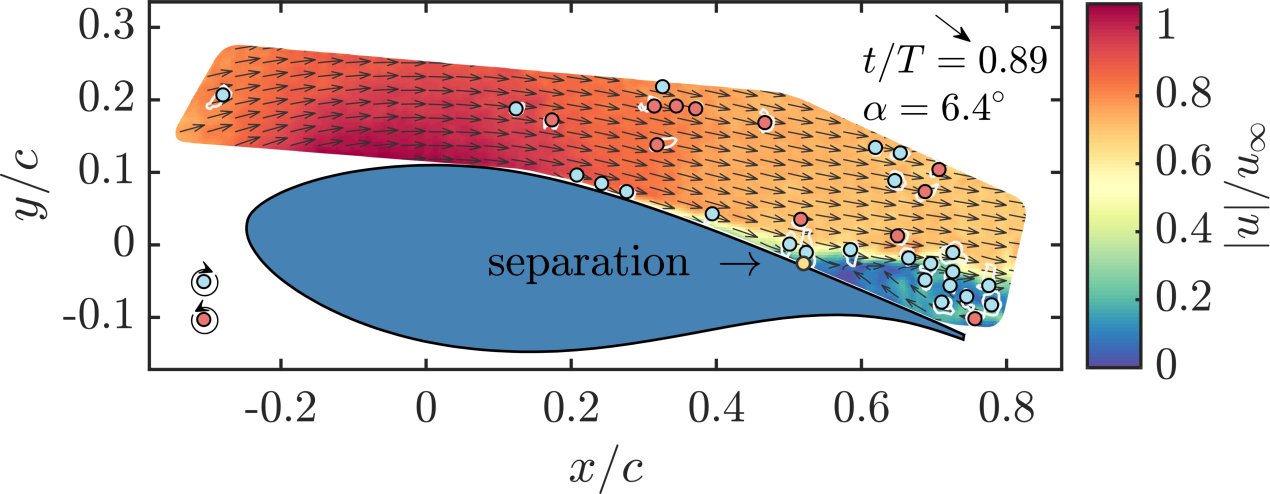}
	}
	
	\subfloat[][Reattachement at $\alpha_{\rm min}$.]{
		\label{fig:2topoReatt-e}
		\includegraphics[height=3.35cm]{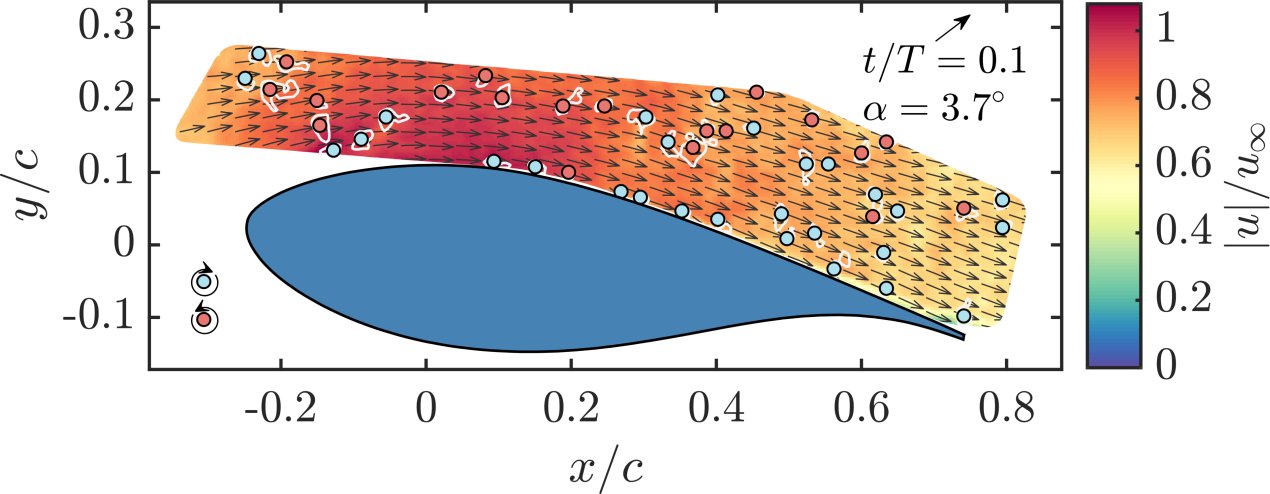}
	}
	\caption[]{Subfigures \subref{fig:2topoAtt-a} -- \subref{fig:2topoReatt-e} show snapshots of five identified characteristic stages of light dynamic stall as a consequence of three-dimensional inflow. $\alpha$ denotes AoA.}
	\label{fig:2topo}
\end{figure}

Five characteristic stages are identified in the present light dynamic stall cycle, shown as snapshots in Figures \ref{fig:2topo}\subref{fig:2topoAtt-a} -- \subref{fig:2topoReatt-e}. The cycle starts during upstroke when the flow is fully attached to the airfoil at $t/T=0.36$, see Figure \ref{fig:2topo}\subref{fig:2topoAtt-a}. The inflow does not show significant perturbations as no vortices are detected in the ambient flow. The instant $t/T=0.36$ corresponds to an AoA of $\alpha\approx10^{\circ}$ according to Figure \ref{fig:1InflowAlphaH}. Such AoA would trigger stall in the static counterpart. Instead, a slight growth of boundary layer towards the leading edge is observed. This comes along with formation of clockwise rotating vortices that are fed by vorticity within the boundary layer once the flow is subjected to an adverse pressure gradient. These vortices follow the envelope of the boundary layer detaching slightly from the surface while they advect downstream.

The onset of light stall is observed in the second stage of the present stall cycle, see Figure \ref{fig:2topo}\subref{fig:2topoOnset-b}. The flow separates close to the trailing edge during the last part of upstroke at $t/T=0.48$. As a consequence of increasing adverse pressure gradient, the boundary layer grows and clockwise rotating vortices emerge frequently along the shear layer. Although the ambient flow contains perturbations that manifest as vortices, the boundary layer vortices remain bound to the surface of the airfoil (see white vortex boundaries). Boundary layer growth is however promoted by inflow perturbations as indicated at $x/c\approx0.4$.

Stage III of the stall cycle is characterized by the maximum extent of the separated flow region, as shown in Figure \ref{fig:2topo}\subref{fig:2topoMax-c}. This happens during downstroke and does not coincide with $\alpha_{\rm max}$. Related aerodynamic details are described in Section \ref{sec:3aero}. The shear layer confining the recirculation region is subjected to secondary instabilities. Shear layer vortices roll-up and vorticity agglomerates in medium-scale structures which move into the recirculation region, thus dividing it into packages. Volatile zones of velocity close to zero form in between those packages. There are counterclockwise rotating vortices during this stage that frequently emerge solely within the separated flow region. Those instants are often related to specific aerodynamic events, as discussed in the following subsection.

Stage IV represents forced reattachment of the separation region by reversing inflow during downstroke. In Figure \ref{fig:2topo}\subref{fig:2topoForced-d}, there is no indication of destabilization of the boundary layer upstream the separation point. The boundary layer then steeply detaches from the surface. The recirculation zone is thus constrained by the ambient flow and forced to advect downstream. The inflow does not contain many vortices. There are however vortices emerging in the ambient flow above the airfoil for $x/c>0.2$ in a region of adverse pressure gradient and an associated increased receptivity for perturbations.

Stage V of the light dynamic stall cycle happens at the lower angular turning point of the inflow at $t/T \approx 0.1$, see Figure \ref{fig:2topo}\subref{fig:2topoReatt-e}. Identified vortices frequently occur in the inflow, as also observed at the upper angular turning point. The surface bound flow is however not affected by ambient instabilities as a consequence of the small AoA of $\alpha=3.7^{\circ}$.

\begin{figure}
	\centering
	\captionsetup[subfigure]{position=top,justification=RaggedRight,singlelinecheck=off}
	\subfloat[][Attached flow during upstroke.]{
		\label{fig:2topoAttZ-a}
		\includegraphics[height=3.35cm]{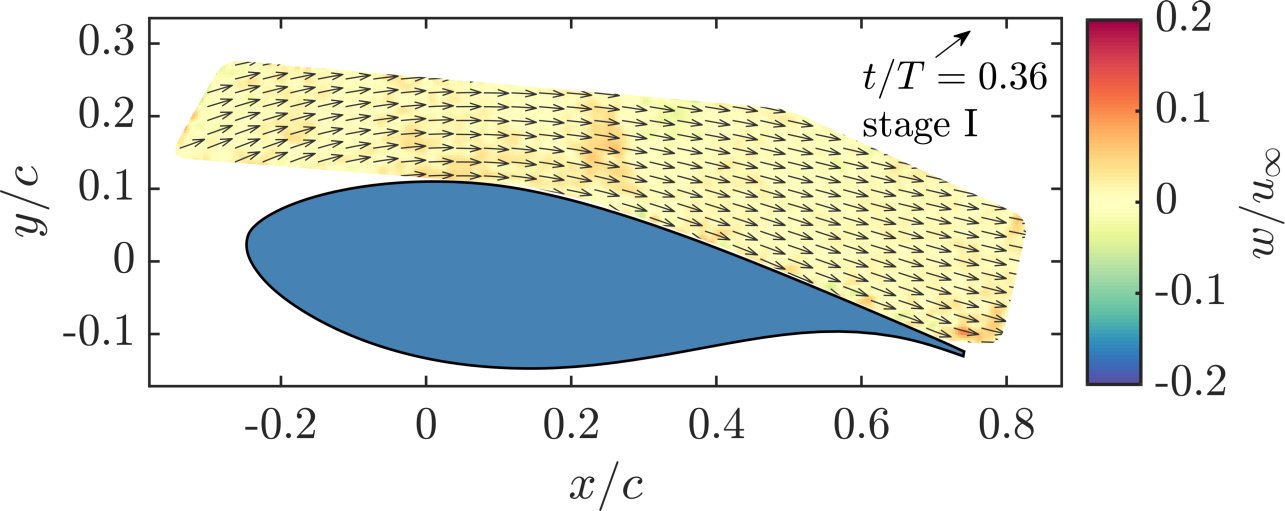}
	}
	
	\subfloat[][Onset of stall during upstroke.]{
		\label{fig:3topoOnsetZ-b}
		\includegraphics[height=3.35cm]{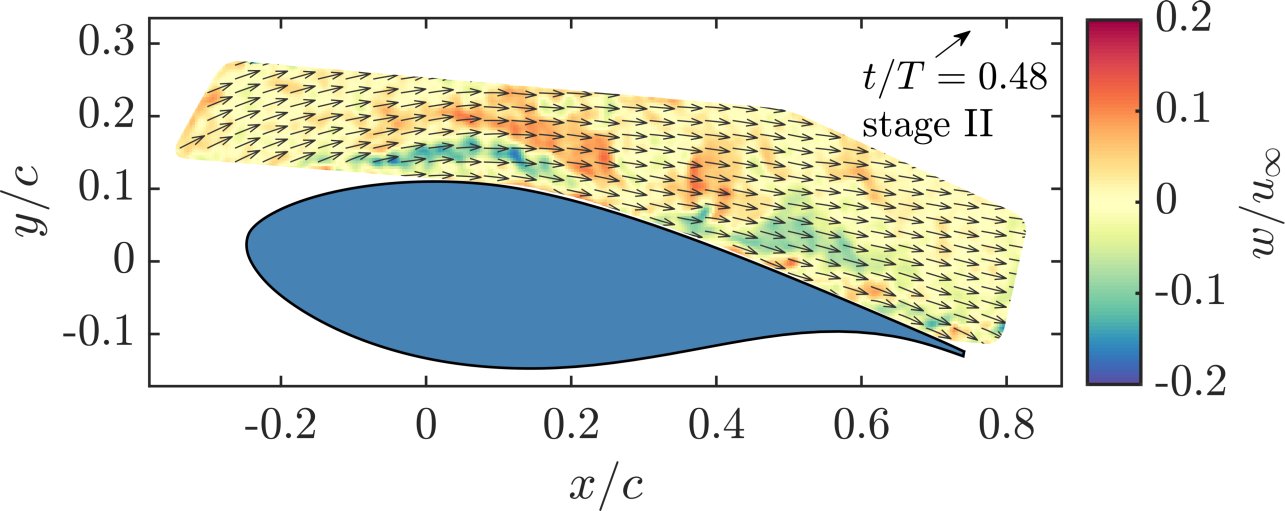}
	}
	
	\subfloat[][Maximum separation after $\alpha_{\rm max}$.]{
		\label{fig:3topoMaxZ-c}
		\includegraphics[height=3.35cm]{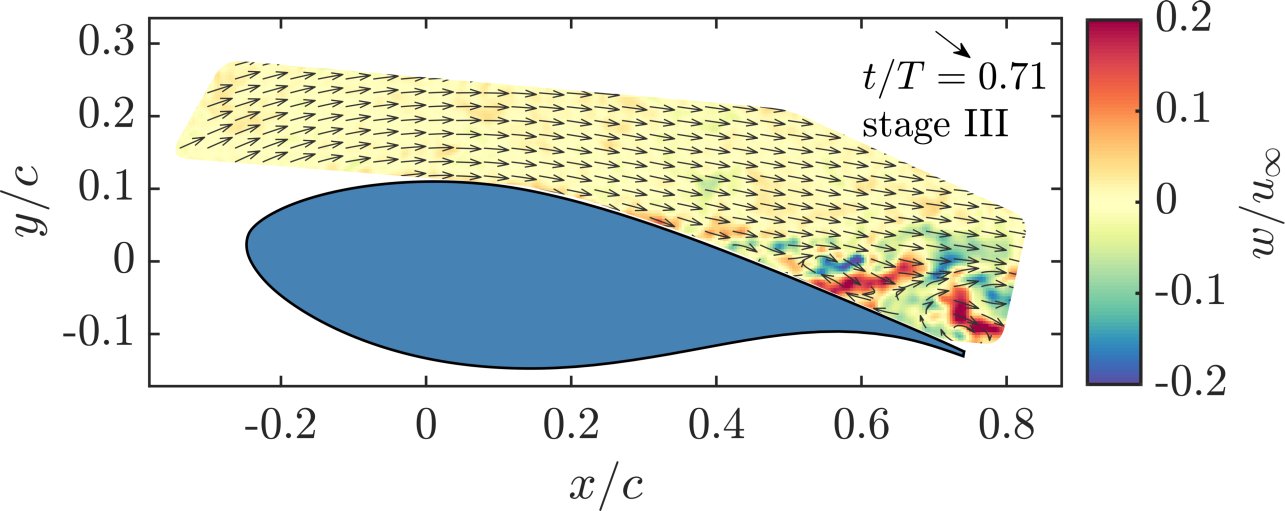}
	}
	
	\subfloat[][Forced shedding of recirculation region during downstroke.]{
		\label{fig:3topoForcedZ-d}
		\includegraphics[height=3.35cm]{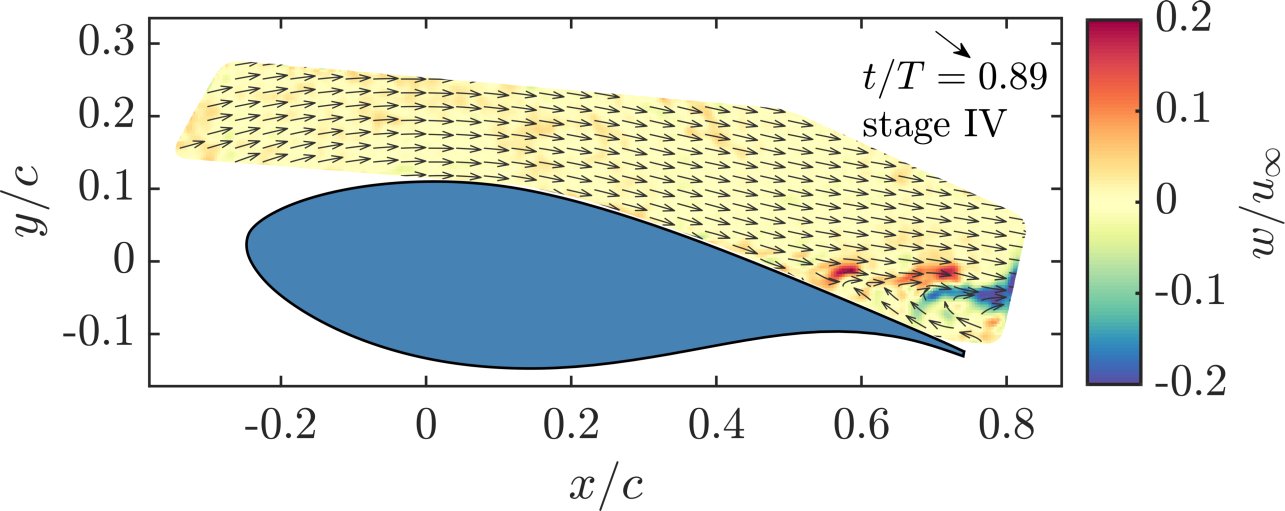}
	}
	
	\subfloat[][Reattachement at $\alpha_{\rm min}$.]{
		\label{fig:3topoReattZ-e}
		\includegraphics[height=3.35cm]{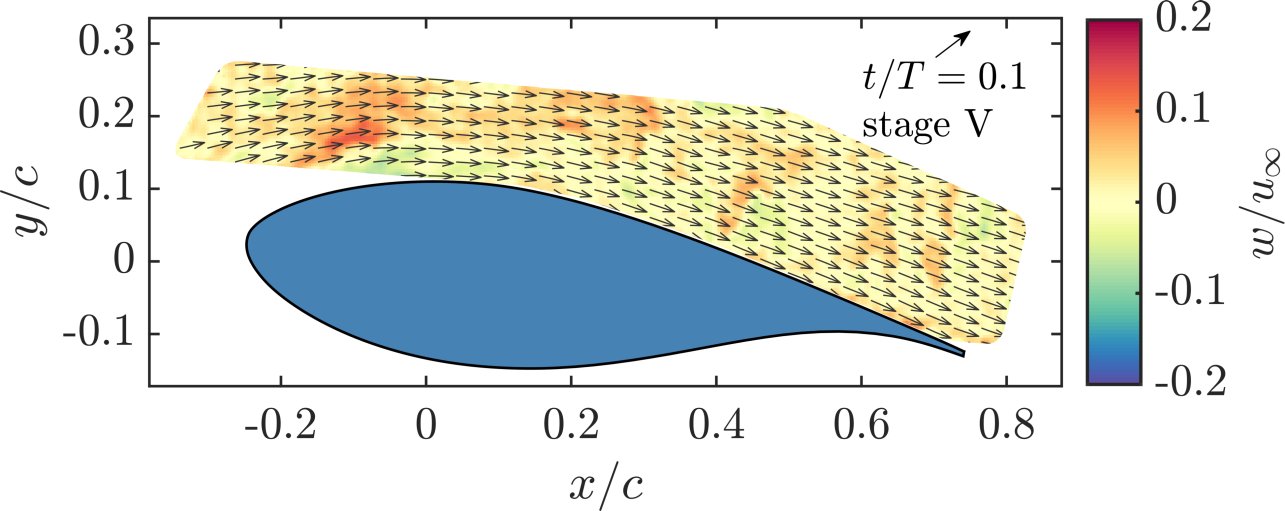}
	}
	\caption[]{Subfigures \subref{fig:2topoAttZ-a} -- \subref{fig:3topoReattZ-e} show velocity fields   of characteristic dynamic stall stages. The spanwise velocity component $w$ is color coded.}
	\label{fig:3topoZ}
\end{figure}

\paragraph{Three-dimensional flow behavior:}
The three-dimensional characteristics of flow formation above the airfoil are analyzed following the identified stall stages. The same instants shown in Figure \ref{fig:2topo} are used in Figure \ref{fig:3topoZ} to investigate the velocity component in spanwise direction, $w$. Due to the relative magnitude with respect to the free stream and chordwise velocities, the spanwise velocity does not play a dominant role over the total cycle. As long as the flow is fully attached to the airfoil during upstroke, negligible spanwise transfer of momentum is observed according to Figure \ref{fig:3topoZ}\subref{fig:2topoAttZ-a}. There is no spanwise flow that is bound to the airfoil and involved in the dynamic parts of the cycle either, i.e. when the separation region emerges or disappears, as shown Figures \ref{fig:3topoZ}\subref{fig:3topoOnsetZ-b} and \subref{fig:3topoReattZ-e}, respectively. The ambient flow however contains spanwise velocity fluctuations of less than $10~\%$ magnitude with respect to the free stream $u_\infty$. On the other hand, stages III and IV reveal significant velocity in spanwise direction of more than $20~\%$ with respect to $u_\infty$ within the recirculation zone, see Figures \ref{fig:3topoZ}\subref{fig:3topoMaxZ-c} and \subref{fig:3topoForcedZ-d}. This corresponds to momentum transfer along the span of the airfoil.

\subsection{Aerodynamic performance of the airfoil}\label{sec:3aero}
\begin{figure}
	\centering
	\includegraphics[width=0.6\linewidth]{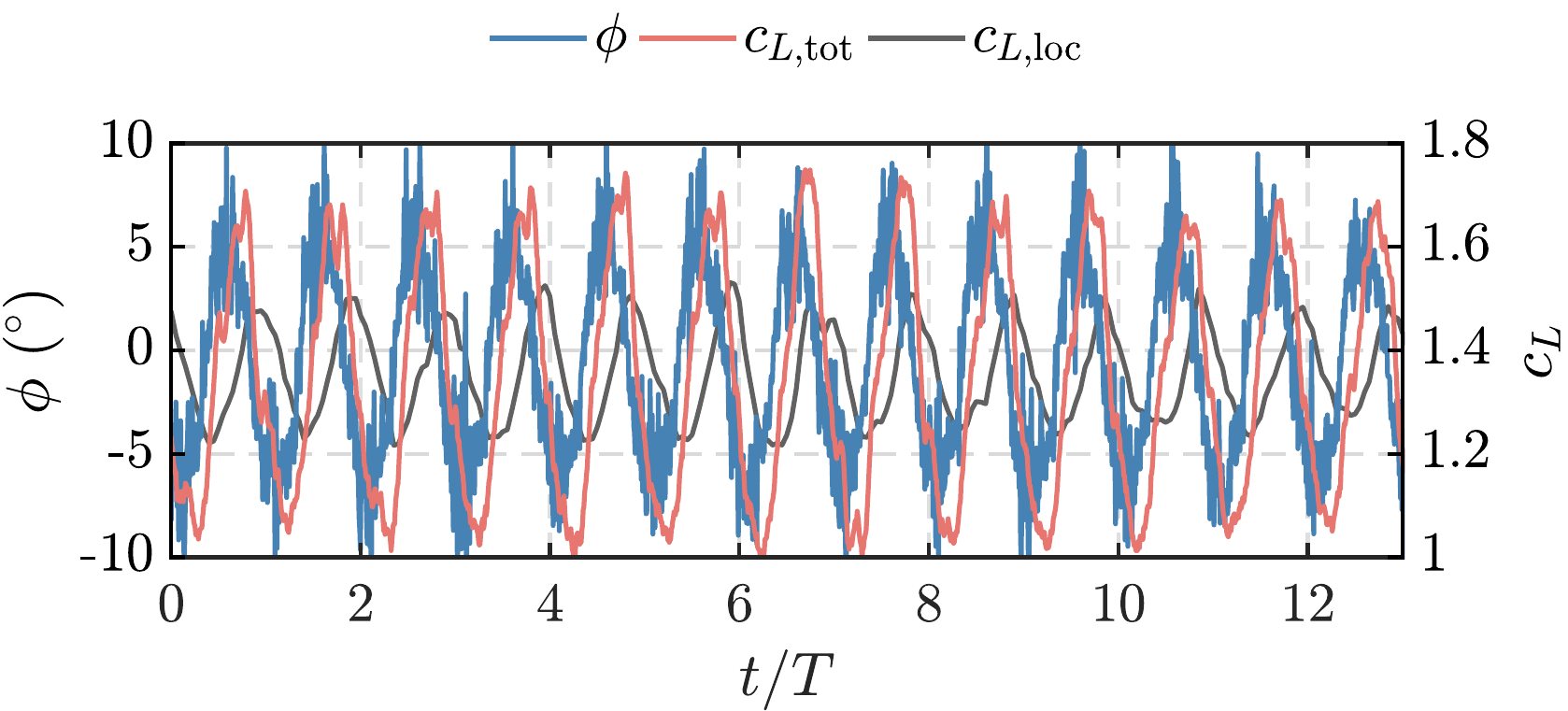}
	\caption{Evolution of AoI $\phi$ compared to lift coefficients $c_{\rm L}$ measured globally (subscript: tot) by means of a force balance and locally (subscript: loc) by means of pressure distribution around the airfoil.}
	\label{fig:3phiCl}
\end{figure}

\paragraph{The cycle of 3D light dynamic stall:}
The flow topology around the airfoil including associated flow phenomena translate into aerodynamic forces. 
The evolution of local and global lift coefficients as well as local AoI are thus shown in Figure \ref{fig:3phiCl}. Local lift coefficient, $c_{\rm L,loc}$, is derived from surface pressure distribution around the airfoil at $z/c=0$ (cf. Figure \ref{fig:0windtunnel}\subref{fig:0windtunnel-a}) and reflects local flow properties. The evolution of $c_{\rm L,loc}$ is smooth because of the connected tubing acting as a low-pass filter. Compared with total lift, $c_{\rm L,tot}$, obtained by the force balance, both values of $c_{\rm L}$ show a $5~{\rm Hz}$ sinusoidal behavior following the inflow. The amplitude of local lift coefficient is almost by $\Delta c_{\rm L}=0.2$ lower than the global one.

\begin{figure}
	\centering
	\includegraphics[width=0.6\linewidth]{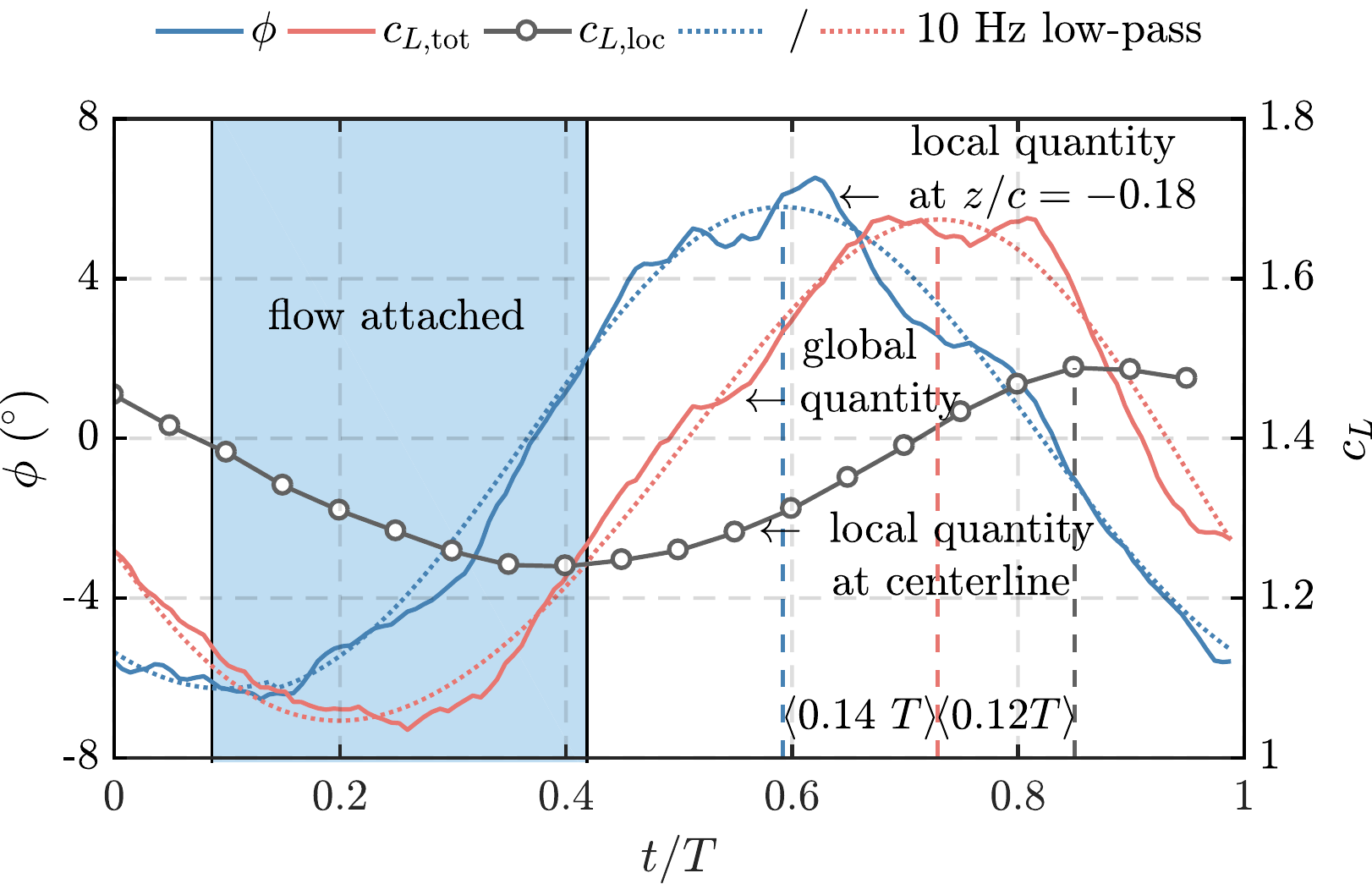}
	\caption{Phase averaged evolution of free stream AoI, $\phi$, without airfoil at quarter chord compared to lift coefficients, $c_{\rm L}$, measured globally by means of a force balance (subscript: tot) and locally by means of pressure distribution around the airfoil (subscript: loc).}
	\label{fig:3phiClphase}
\end{figure}

In quasi-steady airfoil aerodynamics, lift predominantly follows AoA linearly as long as flow remains attached. Comparing the development of AoA (here expressed by AoI), the local and total lift coefficients in Figure \ref{fig:3phiCl}, all three show a difference in their respective phase: AoA peaks first, then the global lift coefficient, followed by the local lift coefficient. The three quantities are shown as a phase average in Figure \ref{fig:3phiClphase} to emphasize this phase relation. The time range of the flow being attached to the airfoil is also highlighted. This is extracted from the time series of snapshots at $z/c=-0.18$ (TRSPIV configuration \#3) in the same manner as described concerning Figure \ref{fig:2detFlow} and phase averaged in time afterwards. The AoI as well as global $c_{\rm L,tot}$ reveal characteristics in their respective development, such as an unclear maximum, that complicates the determination of a phase. Both signals are thus at first low-pass filtered at 10 Hz and then phase-averaged. By comparing their maxima with the maximum of local $c_{\rm L,loc}$ in Figure \ref{fig:3phiClphase}, phase shifts of $0.14~T$ between local AoI and global lift coefficient and $0.12~T$ between global and local lift coefficient are obtained. These phase shifts are considered as equal with respect to the temporal resolution of local $c_{\rm L}$ at the centerline. Global lift maximum thus arrives temporally in the middle of the two local quantities, AoI and $c_{\rm L,loc}$.

\begin{figure}
	\centering
	\includegraphics[width=0.6\linewidth]{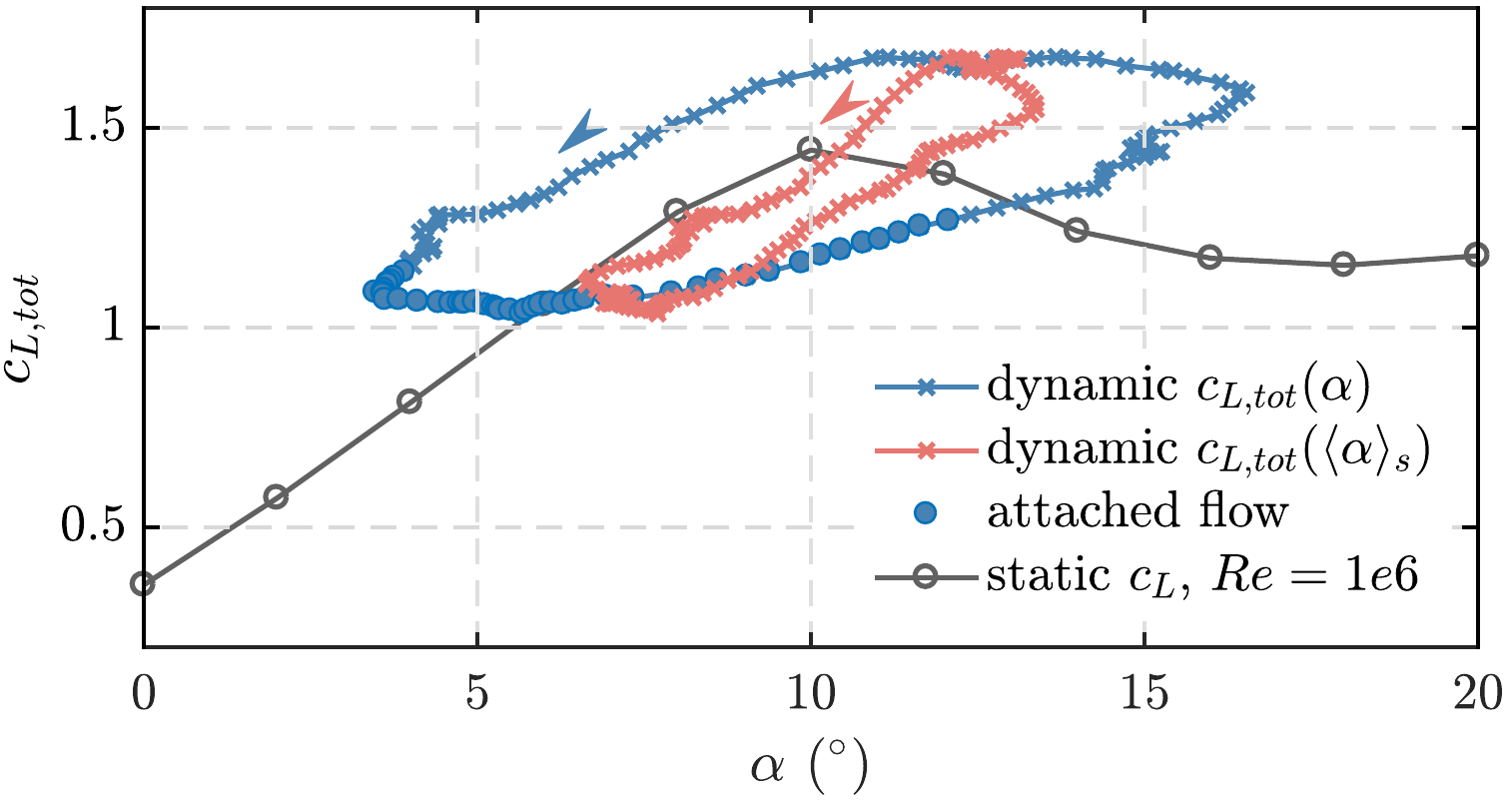}
	\caption{Dynamic polars consisting of the local AoA, $\alpha$, at $z/c=-0.18$, mean AoA along the span of the airfoil, $\langle \alpha \rangle_s$, and the total lift coefficient, $c_{\rm L, tot}$ (blue line), are compared to static lift, $c_{\rm L}$. The arrows indicate the direction of cyclic evolution. Aerodynamic states of attached flow are emphasized by filled symbols.}
	\label{fig:2alphaCl}
\end{figure}

The phase relationship of AoA and lift is often presented as a dynamic polar in studies on dynamic stall \cite{Carr1988}. Such depiction forms characteristic closed loops that give deeper insight into flow effects and experimental conditions. In Figure \ref{fig:2alphaCl}, the total lift coefficient is plotted as a function of AoA in a phase-averaged manner. Filled symbols mark attached flow and the arrow points into the direction of temporal evolution. The static lift curve is also shown for comparison. The flow is attached to the airfoil at $\alpha_{\rm min}$ in the dynamic case. As AoA increases, $c_L$ starts with a value above the static polar but increases with a smaller slope. In quasi-two-dimensional dynamic stall, these effects are usually associated with a delayed response of the flow around the airfoil yielding an apparently thicker airfoil along with increased apparent camber. Both effects are well known to decrease the lift slope and shift the lift curve upward \cite{Choudhry2014}, as also seen in Figure \ref{fig:2alphaCl}. Dynamic stall then takes effect by delaying stall until $\alpha = 12^\circ$ and an overshoot in lift for $10^\circ < \alpha \lesssim 16^\circ$.

In most dynamic stall cases, where the airfoil undergoes either light or deep stall and experiences low or high reduced frequency dynamics, dynamic stall cycles proceed clockwise due to a sudden drop in lift while the AoA follows behind (negative phase shift) \cite{McCroskey1982,Carr1988}. The present cycle however appears counterclockwise which means that the phase shift between $\alpha$ and $c_{\rm L}$ is positive. The downstroke of $\alpha$ still augments lift to $c_{\rm L} \approx 1.7$, which is $16~\%$ above static lift. Dynamic lift remains above the static lift curve during the full downstroke of $\alpha$. Since the evolution of AoA refers to a local measurement but lift is a global quantity, the dynamic polar with respect to the spatial mean AoA, $\langle \alpha \rangle_s$, is also shown in Figure \ref{fig:2alphaCl}. The mean AoA facing the airfoil is obtained by averaging AoA in spanwise $z$-direction taken from data presented in Figure \ref{fig:1InflowAlphaV}. The AoA range decreases due to spatial averaging and a part of the downstroke thus takes place below the static lift curve. The dynamic polar is still directed counterclockwise. This result is discussed in Section \ref{sec:5disc}.

\begin{figure*}
	\centering
	\captionsetup[subfigure]{position=top,justification=RaggedRight,singlelinecheck=off}
	\subfloat[][]{
		\label{fig:2vidCVa-a}
		\includegraphics[width=0.5\linewidth]{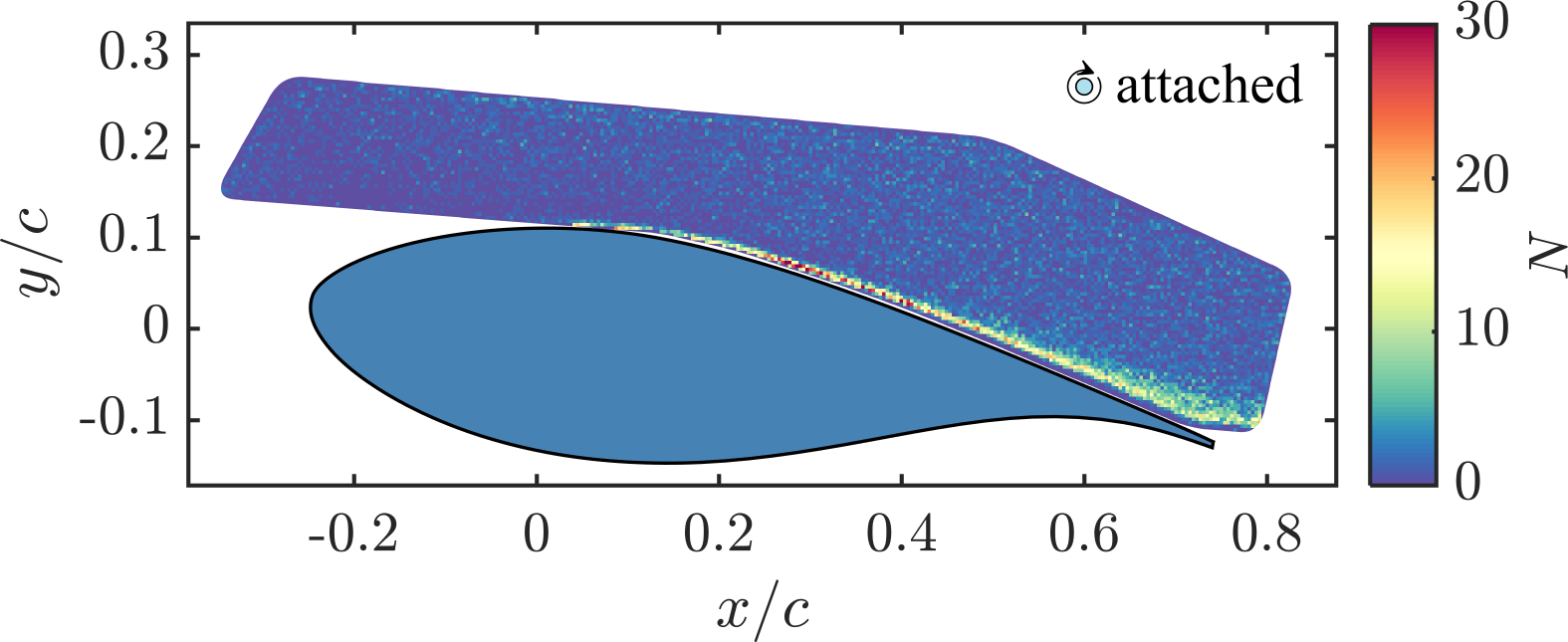}
	}
	\subfloat[][]{
		\label{fig:2vidCCVa-b}
		\includegraphics[width=0.5\linewidth]{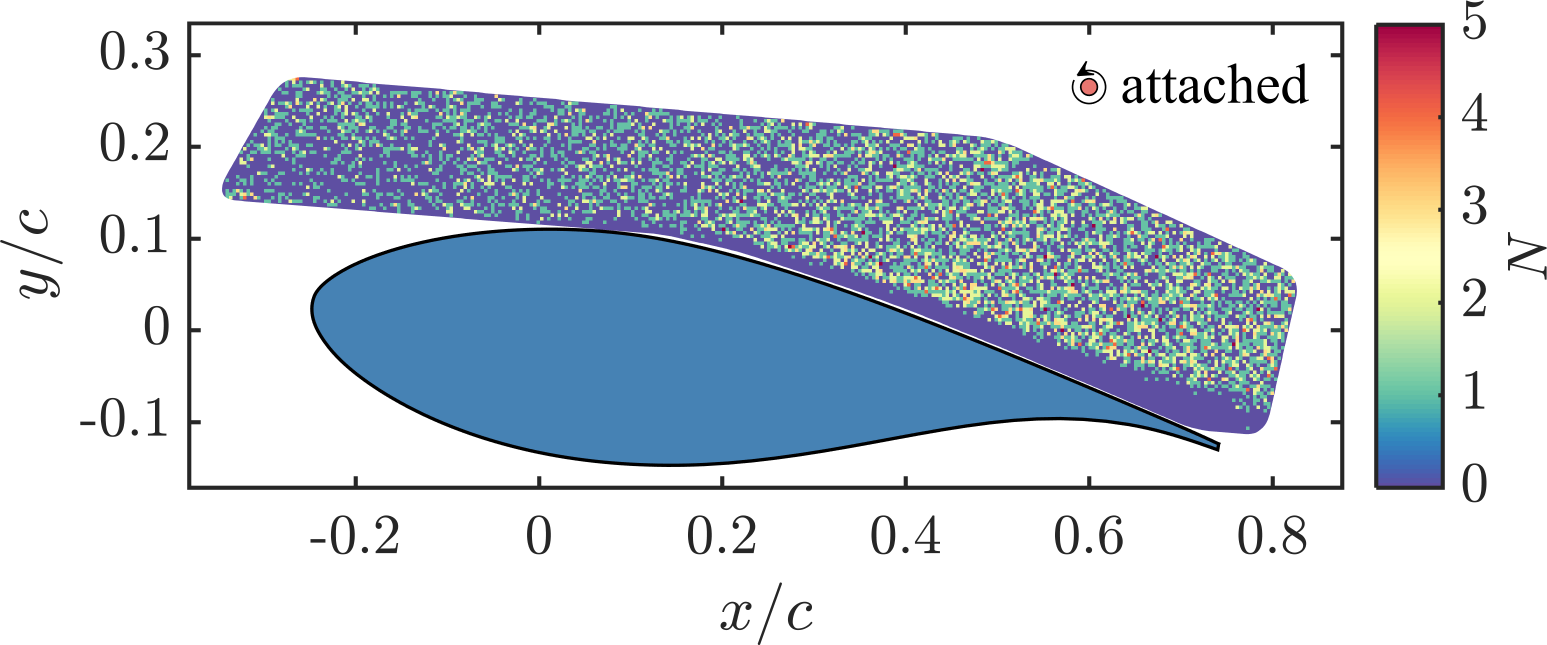}
	}
	
	\subfloat[][]{
		\label{fig:2vidCVs-c}
		\includegraphics[width=0.5\linewidth]{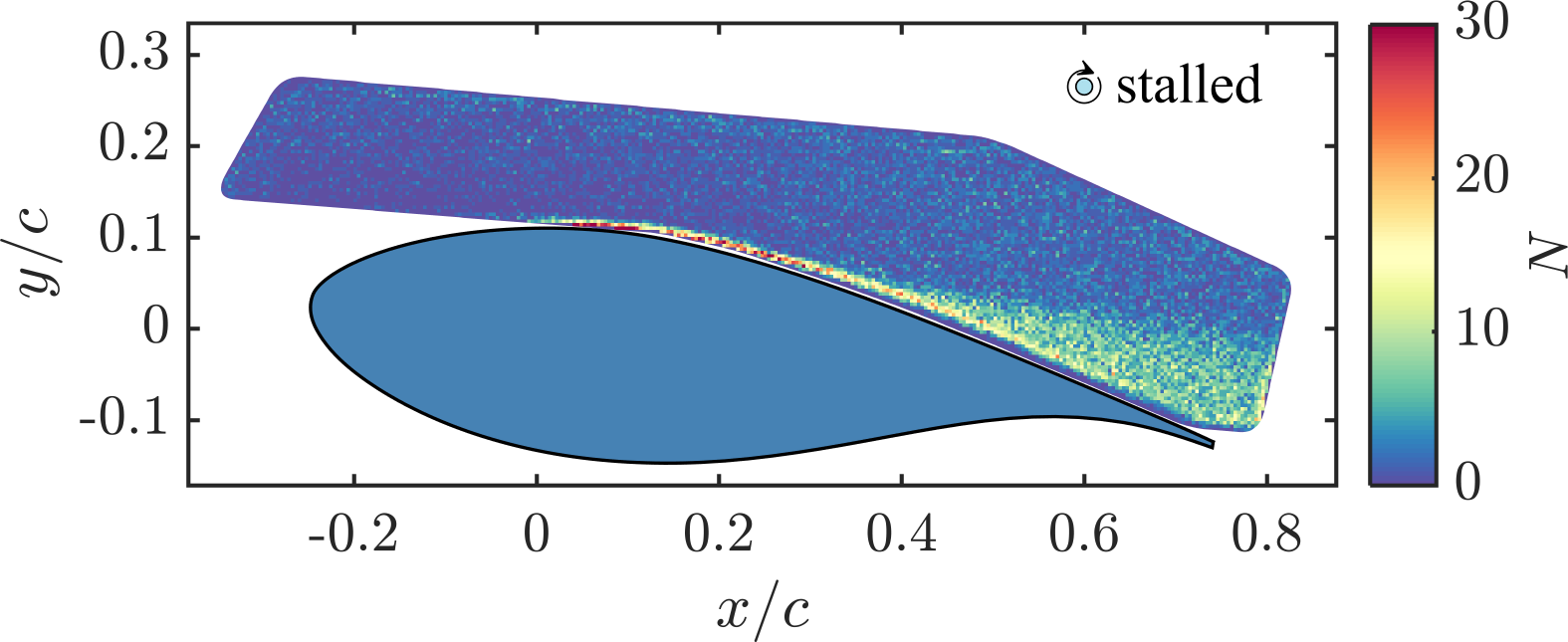}
	}
	\subfloat[][]{
		\label{fig:2vidCCVs-d}
		\includegraphics[width=0.5\linewidth]{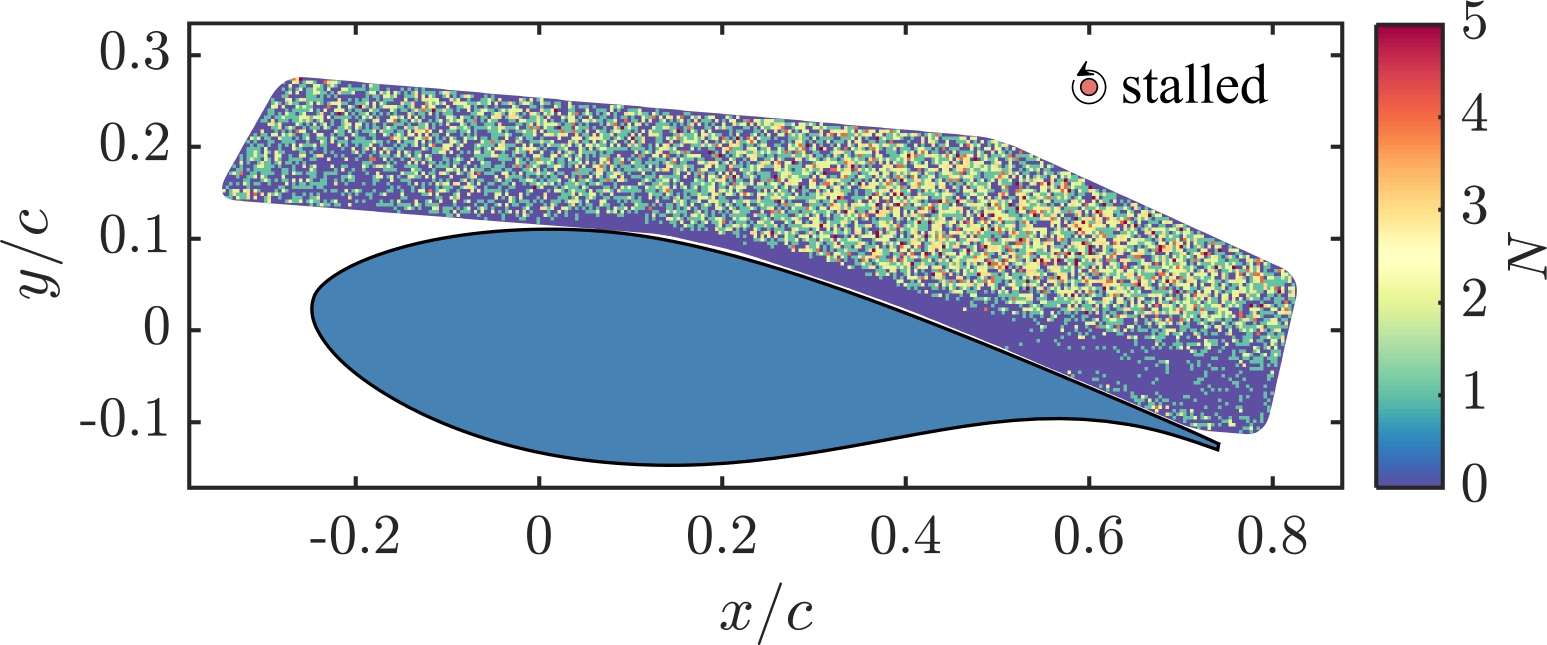}
	}
	\caption[]{Spatial histogram of clockwise (\subref{fig:2vidCVa-a}, \subref{fig:2vidCVs-c}) and counterclockwise (\subref{fig:2vidCCVa-b}, \subref{fig:2vidCCVs-d}) rotating vortices while flow is attached (\subref{fig:2vidCVa-a}, \subref{fig:2vidCCVa-b}) and stalled (\subref{fig:2vidCVs-c}, \subref{fig:2vidCCVs-d}) which is distinguished by the presence of reversed flow.}
	\label{fig:2vid}
\end{figure*}

\paragraph{Vortex airfoil interaction:}
The present flow topology, ranging from inflow to the formation of a recirculation zone above the airfoil, is complex and multi-scale consisting of an interaction of numerous vortices, as seen in Figures \ref{fig:2detFlow} and \ref{fig:2topo}. To shed light onto such interaction, the spatial distribution of emerging vortices about the airfoil is determined. Since the attached and stalled stages are fundamentally different, statistics of counterclockwise/clockwise rotating vortices are conditioned on stalled/attached flow state and shown in Figures \ref{fig:2vid}\subref{fig:2vidCVa-a}--\subref{fig:2vidCCVs-d}. The number of vortex observations at a certain location is color-coded giving a two-dimensional histogram. This representation also indicates mean trajectories of vortices. Starting with the attached stage in Figure \ref{fig:2vid}\subref{fig:2vidCVa-a}, clockwise rotating vortices significantly detach from the airfoil quite downstream at $x/c\approx0.4$ and in a very localized manner. In contrast, during stalled stage in Figure \ref{fig:2vid}\subref{fig:2vidCVs-c}, detachment occurs at $x/c\approx0.3$ together with clockwise rotating vortices strongly spreading in $y$-direction. This property serves as an identification of light dynamic stall \cite{Mulleners2012}. In both cases, downstream advection of vortices goes along with growing distance to the surface of the airfoil as a consequence of the formation of recirculation as well as rotation direction and the associated low pressure region above the respective vortex. Low relative density of clockwise rotating vortices (Figure \ref{fig:2vid}\subref{fig:2vidCVa-a}) visually complements the region of high relative density of counterclockwise rotating vortices, which mainly occur within the ambient flow (Figure \ref{fig:2vid}\subref{fig:2vidCCVa-b}). In the stalled case shown in Figure \ref{fig:2vid}\subref{fig:2vidCCVs-d}, this looks very much alike apart from one remarkable difference that is revealed close to the trailing edge in the very vicinity of the airfoil. There is an agglomeration of counterclockwise rotating vortices observed.

\begin{figure*}
	\centering
	\captionsetup[subfigure]{position=top,justification=RaggedRight,singlelinecheck=off}
	\subfloat[][]{
		\label{fig:3topoCCV_III-a}
		\includegraphics[width=0.5\linewidth]{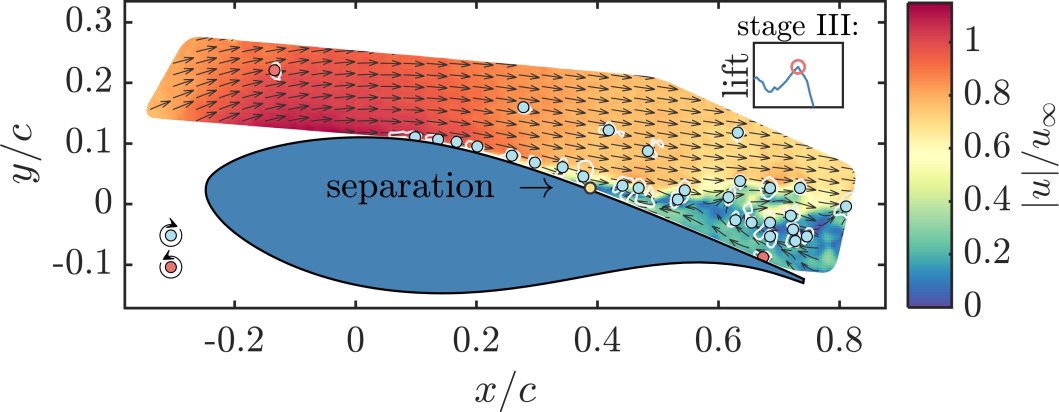}
	}	
	\subfloat[][]{
		\label{fig:3topoCCV_III-b}
		\includegraphics[width=0.5\linewidth]{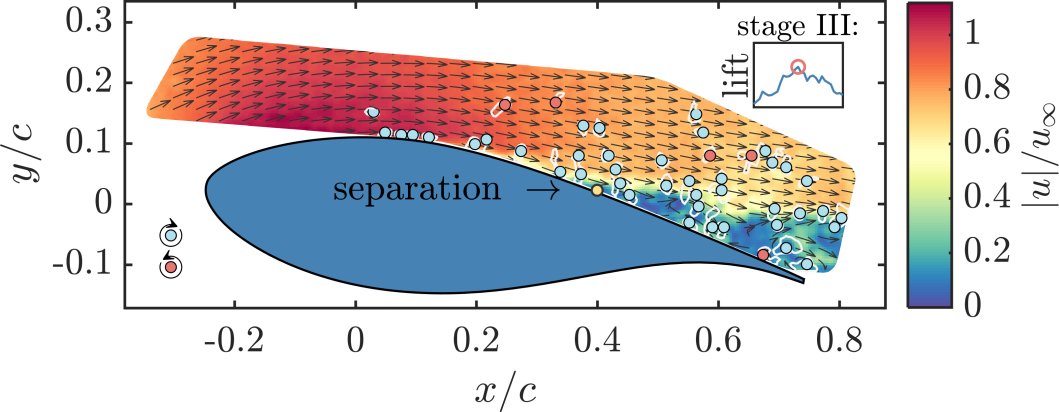}
	}
	\caption[]{Comparison of velocity fields occuring in stage III with instantaneous lift (insert) in Subfigures \subref{fig:3topoCCV_III-a} -- \subref{fig:3topoCCV_III-b}. In many cases, locally emerging counterclockwise rotating vortices coincide with remarkable global lift events.}
	\label{fig:3topoCCV_III}
\end{figure*}

A closer look is thus taken into the flow field of those instances where counterclockwise rotating vortices emerge within the recirculation region. A remarkable coincidence becomes apparent when each snapshot of stalled flow and the instantaneous lift behavior are considered. In 24 cases, from which a selection of two is shown in Figures \ref{fig:3topoCCV_III}\subref{fig:3topoCCV_III-a} and \subref{fig:3topoCCV_III-b}, during all light dynamic stall cycles, the emergence of one single counterclockwise rotating vortex is associated with a relative maximum in global lift. For clarification, a detail of lift evolution is inserted into the upper right corner of each subfigure together with the corresponding lift event marked by a red circle. These extraordinary instances emerge in stage III of the light dynamic stall cycle only. How this observation relates to counterclockwise dynamic stall cycle and the three-dimensional flow formation is discussed in Section \ref{sec:5disc}.

\subsection{Aerodynamic footprint by means of POD}\label{sec:4pod}
The POD is a statistical method that has proven to be useful for extracting relevant features from complex flow situations in aerodynamics \cite{Melius2016,Traphan2018}. A snapshot POD of an associated data set of velocity field yields a set of normalized spatial eigenmodes, $\Phi_{\rm i}$, of the total time series. They represent spatial coherence ranked by its respective content of turbulence kinetic energy that is captured by the corresponding eigenvalue, $\lambda_{\rm i}$. Each spatial eigenmode is paired with a temporal weighting coefficient, $a_{\rm i}(t)$. The contribution of a spatial eigenmode to a velocity field at a certain time, $t$, is thus given by the magnitude of its weighting \cite{Bernero2000}.

Since the presented flow about the airfoil is unsteady as well as periodic, POD needs a certain number of cycles to converge until variations between cycles are captured. For verifying whether 13 measured stall cycles are sufficient, 13 PODs are briefly compared, each includes a different number of inflow cycles for computation. As an integrated measure for the convergence of a POD eigenmode, its respective eigenvalue $\lambda_{\rm i}$ is taken. 
The difference between respective eigenvalues is less than $0.3~\%$ of the standard deviation, $\sigma(\lambda_{\rm i})$, after the 6th cycle. This result is in accordance with other studies which found convergence after five cycles for quasi-2D periodic inflow conditions \cite{Mariappan2014}. An analysis of the available 13 cycles is thus sufficient to capture the periodicity of the present flow by means of POD.

\begin{figure*}
	\captionsetup[subfigure]{position=top,justification=raggedright,singlelinecheck=off}
	\centering
	\subfloat[][]{
		\label{fig:4spatEM0}
		\includegraphics[height=3.0cm]{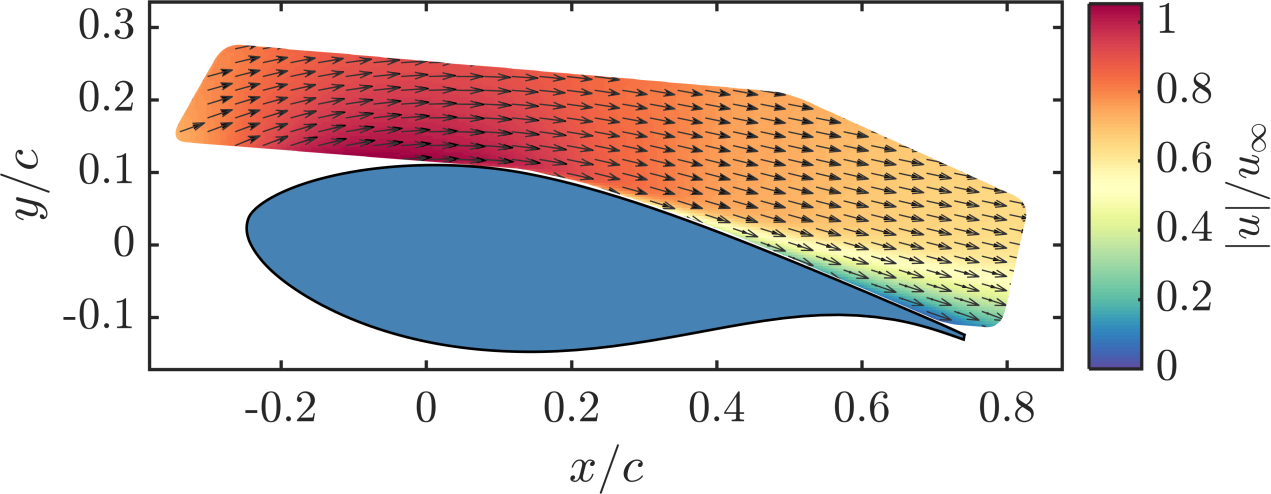}
	}
	~
	\subfloat[][]{
		\label{fig:4spatEM1}
		\includegraphics[height=3.0cm]{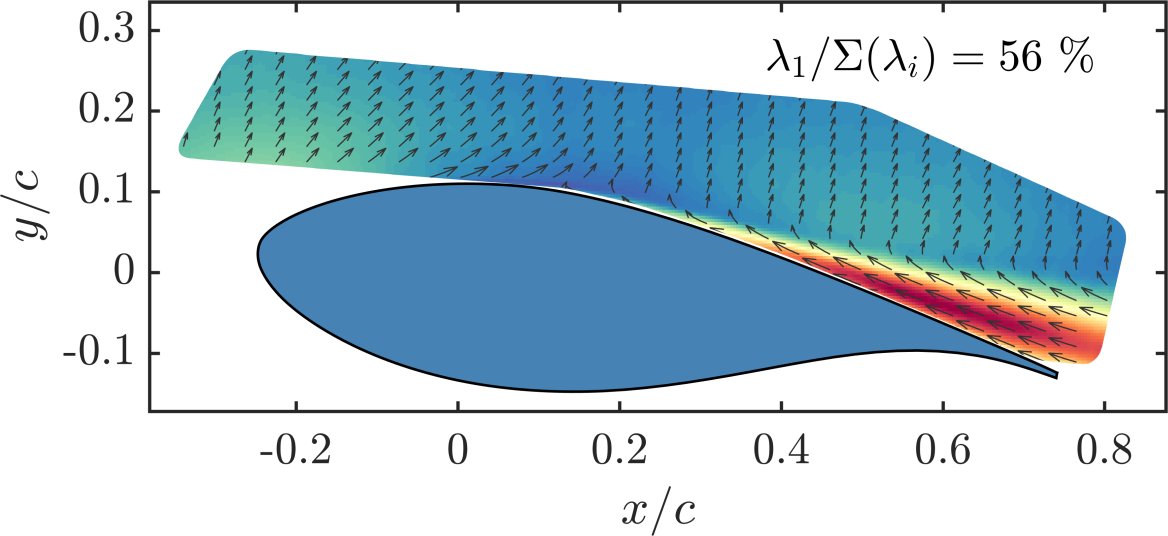}
	}
	
	\subfloat[][]{
		\label{fig:4spatEM2}
		\includegraphics[height=3.0cm]{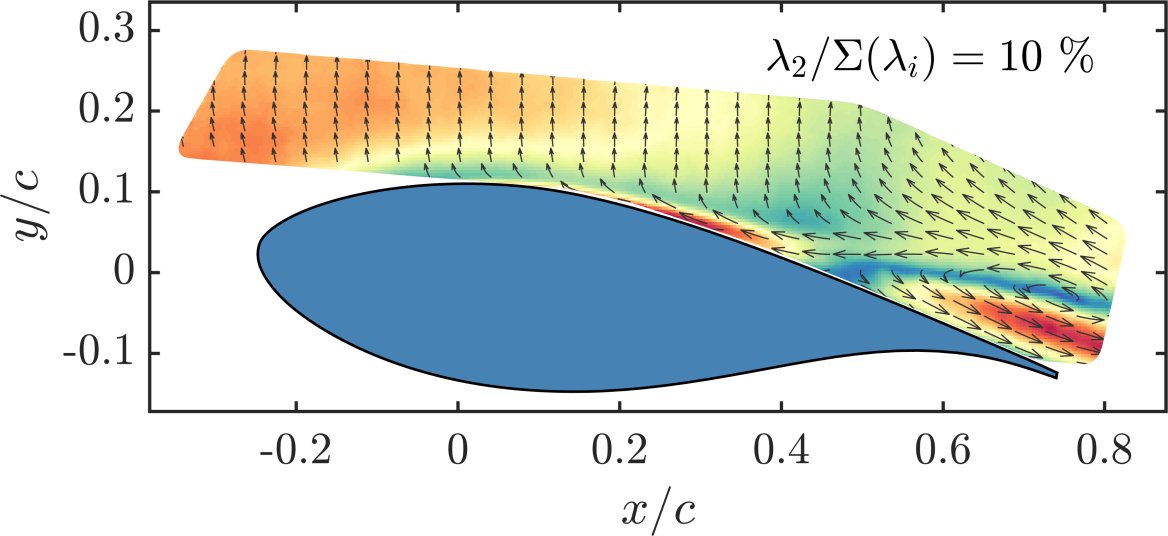}
	}
	~~~~~~~~~~~~
	\subfloat[][]{
		\label{fig:4spatEM3}
		\includegraphics[height=3.0cm]{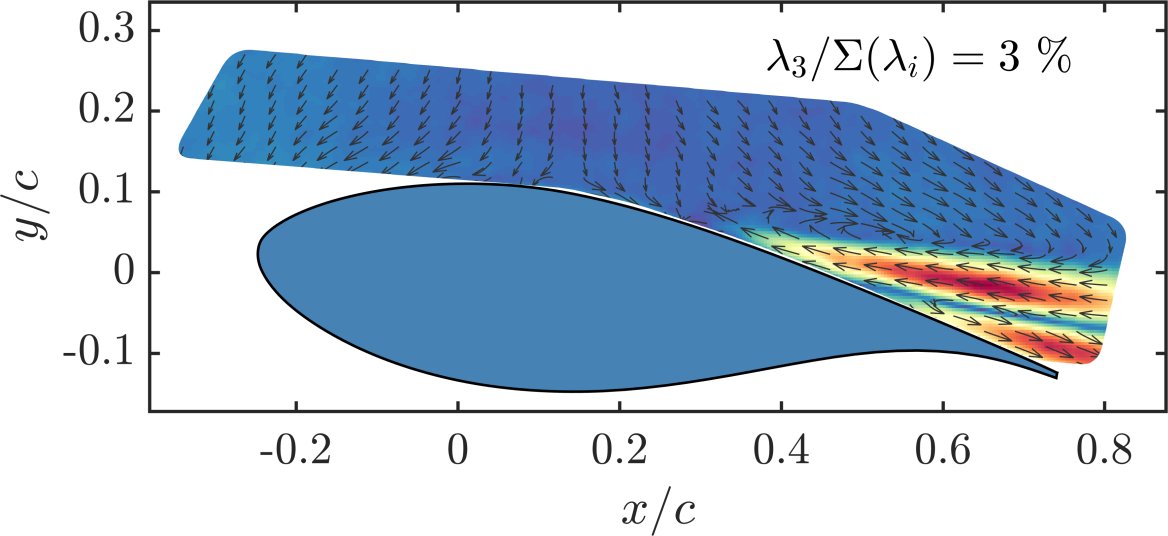}
	}
	\caption[]{Mean velocity field $\langle\vec{u}\rangle$ in \subref{fig:4spatEM0} serves as a basis for POD interpretation in the sense of superposition. The three strongest spatial eigenmodes of velocity field evolution $\Phi_1$--$\Phi_3$ ranked by eigenvalues representing fluctuating energy content are shown in \subref{fig:4spatEM1}--\subref{fig:4spatEM3}.}
	\label{fig:4POD}
\end{figure*}

The presented POD is derived from snapshots of velocity fluctuations $\vec{u}^{\, \prime}(t/T)=\oldvec{u}(t/T)-\langle\oldvec{u}\rangle_t$. Spatial eigenmodes thus reflect the distribution of turbulence kinetic energy of the time series. In Figure \ref{fig:4POD}, the mean flow, $\langle\vec{u}\rangle_t$, as well as the three most energetic spatial eigenmodes, $\Phi_{\rm i}$, with ${\rm i} \in [1, 2, 3]$, are shown. Vector arrows indicate the flow contribution in $x$ and $y$-directions. Each contour represents the normalized velocity contribution of the respective eigenmode (Figures \ref{fig:4POD}\subref{fig:4spatEM1}--\subref{fig:4spatEM3}). The three strongest eigenmodes contribute more than 2/3 to the energy contained within $\vec{u}^{\, \prime}$, i.e. $(\lambda_{\rm 1} + \lambda_{\rm 2} + \lambda_{\rm 3}) / \sum(\lambda_{\rm i}) \approx 69~\%$. Eigenmodes above the 3$^{rd}$ one often consist of higher spatial harmonics of eigenmodes 1--3 and the corresponding energy remains below $1~\%$.

The first spatial POD eigenmode, $\Phi_1$, reveals that $56~\%$ of the total turbulence kinetic energy is focused on the wake region of the airfoil that emerges during stall. In the sense of superposition, the first eigenmode represents the extent of the wake and thus indicates that separation takes place downstream of $x/~c=0.3$. 
This is as expected from the DU airfoil family where deep-stall happens typically for AoA above $20^{\circ}$ \cite{Brondsted2013}. 

The inflow is mainly captured by $\Phi_2$. Almost vertically pointing velocity vectors within the approaching flow modulate the mean flow (Figure \ref{fig:4POD}\subref{fig:4spatEM0}). There are also contributions to the region of adverse pressure-gradient in the vicinity of the airfoil where boundary-layer destabilization happens when AoA increases. In addition to the total stall region represented by $\Phi_1$, $\Phi_2$ contributes significant turbulence kinetic energy to the wake in the region close to the trailing edge. The second eigenmode captures $10~\%$ of the total turbulence kinetic energy.

A drop in energy contribution to $3~\%$ of mode $\Phi_3$ (see Figure \ref{fig:4POD}\subref{fig:4spatEM3}) implies only small scale fluctuations or rare but energetic events are covered. The third POD mode affects a greater region about the stall region than $\Phi_1$. $\Phi_3$ is characterized by flow that is oppositely directed at the shear-layer compared to the vicinity of the airfoil. Thus, contribution of $\Phi_3$ results in deformation of the recirculation zone given by $\Phi_1$.

\begin{figure}
	\centering
	\captionsetup[subfigure]{position=top,justification=RaggedRight,singlelinecheck=off}
	\subfloat[][]{
		\label{fig:4tempEM1-a}
		\includegraphics[width=0.6\linewidth]{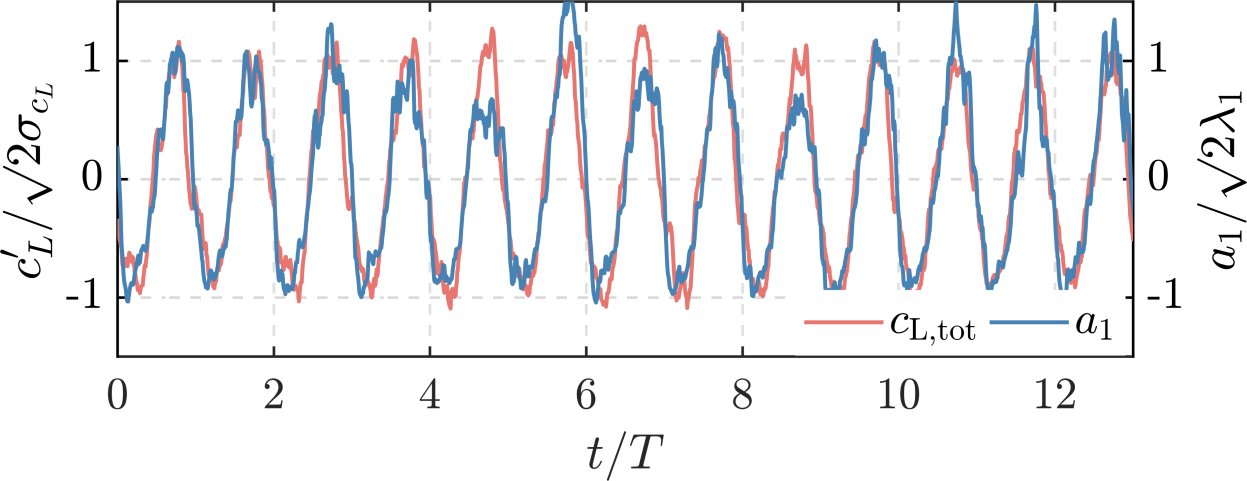}
	}
	\hspace{8pt}
	\subfloat[][]{
		\label{fig:4tempEM1-b}
		\includegraphics[width=0.6\linewidth]{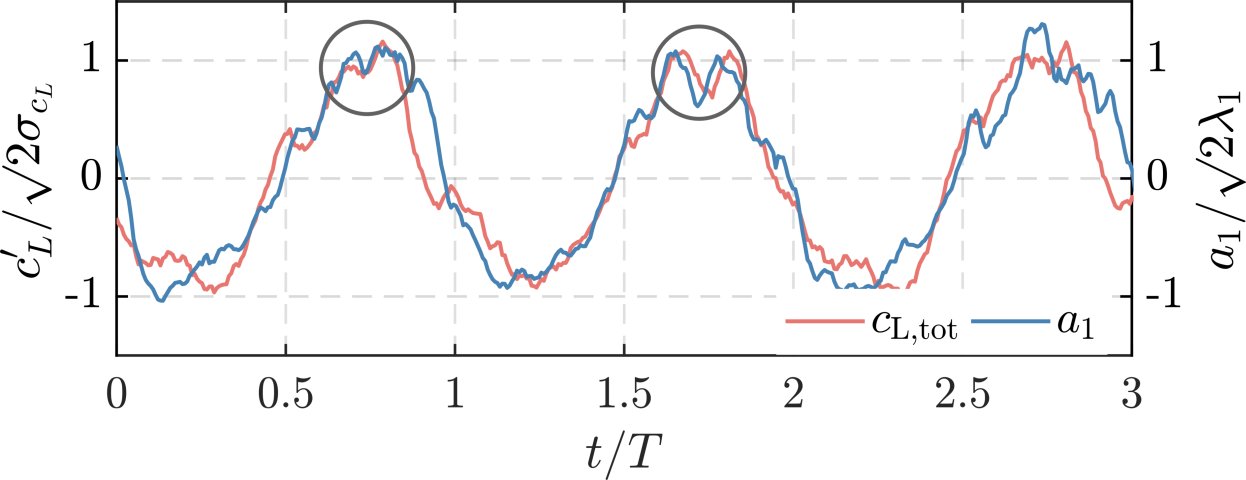}
	}
	\caption[]{\subref{fig:4tempEM1-a} Evolution of first temporal eigenmode of velocity field, ranked by corresponding eigenvalues, in comparison to $c_{\rm L}$. \subref{fig:4tempEM1-b} Magnification of \subref{fig:4tempEM1-a} in order to emphasize coinciding events marked by circles.}
	\label{fig:4PODtemp}
\end{figure}

The temporal contribution of spatial eigenmodes is expressed by the time development coefficients, $a_{\rm i}(t/T)$. In Figure \ref{fig:4PODtemp}, the coefficient $a_{\rm 1}(t/T)$, which is associated with $\Phi_1$, is compared with the evolution of the global lift coefficient in terms of fluctuations denoted by $^\prime$. Both coefficients, $a_{\rm 1}(t/T)$ and $c_{\rm L,tot}$, evolve in phase ($\rho \approx 0.93$) in the expected periodic manner (see Figure \ref{fig:4PODtemp}\subref{fig:4tempEM1-a}). The high lift intervals demonstrate several instances throughout the time series where $a_{\rm 1}(t/T)$ also reflects details, as marked exemplary by circles in Figure \ref{fig:4PODtemp}\subref{fig:4tempEM1-b}. A time lag of $\Delta t/T \approx 0.04$ between those details is observed in some cases, as revealed in the second circled event. 

\begin{figure}
	\centering
	\includegraphics[width=0.6\linewidth]{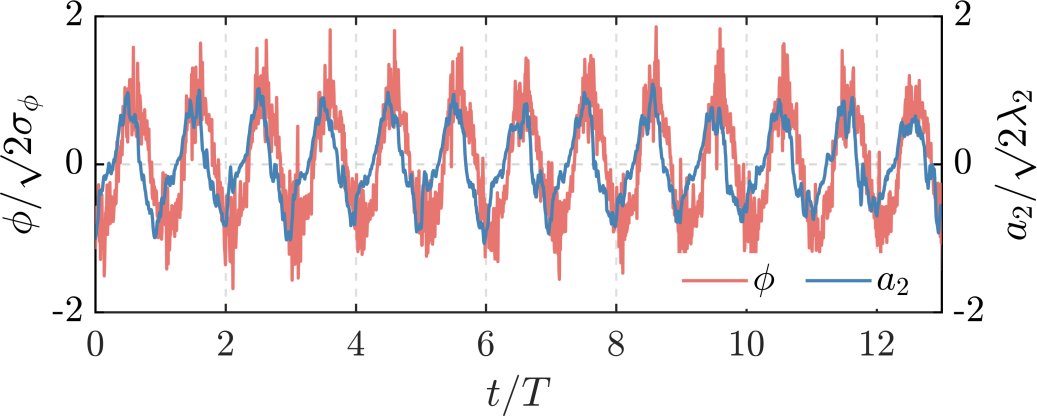}
	\caption{Evolution of second temporal eigenmode of velocity field, ranked by corresponding eigenvalues, in comparison to AoI $\phi$.}
	\label{fig:4tempEM2}
\end{figure}

Since $\Phi_2$ contains major information about the inflow, its temporal contribution is compared with the AoI in Figure \ref{fig:4tempEM2}. Coefficient $a_{\rm 2}$ also shows periodic behavior but leads AoI by $\Delta t/T \approx 0.09$. Some portions of the inflow are thus contributed by other eigenmodes, as also indicated by $\Phi_1$. A comparison of $a_{\rm 2}(t/T)$ and $a_{\rm 1}(t/T)$ yields a phase shift of $\Delta t/T\approx0.14$, which corresponds to the phase shift between AoI and $c_{\rm L,tot}$ (cf. Figure \ref{fig:3phiClphase}).

\begin{figure}
	\centering
	\captionsetup[subfigure]{position=top,justification=RaggedRight,singlelinecheck=off}
	\subfloat[][]{
		\label{fig:4tempEM3}
		~~~~~~\includegraphics[height=3.25cm]{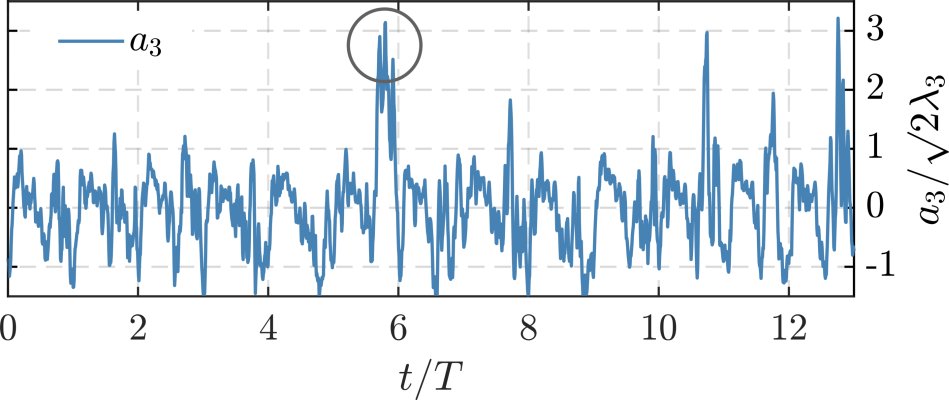}
	}
	
	\subfloat[][]{
		\label{fig:4SSmaxEM3-b}
		\includegraphics[width=0.6\linewidth]{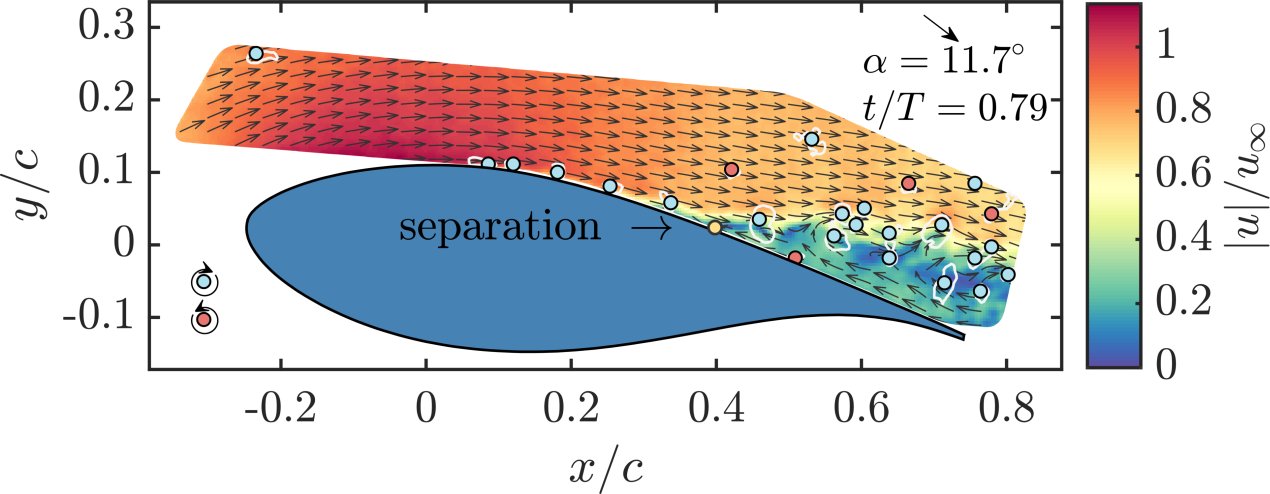}
	}
	\caption[]{\subref{fig:4tempEM3} Evolution of third temporal eigenmode of velocity field, ranked by corresponding eigenvalues. Strong amplitudes of $a_3$, such as at $t/T=5.8$, point towards extraordinary events occurring within the flow. \subref{fig:4SSmaxEM3-b} Corresponding velocity field is depicted including clockwise rotating vortices and counter clockwise rotating ones. During dynamic downstroke, AoA decreases suddenly and the flow is pushed back towards the surface of the airfoil forming a unique flow structure.
	}
	\label{fig:4EM3}
\end{figure}

As a last aspect of POD analysis, the third time development coefficient $a_{\rm 3}(t/T)$ is shown in Figure \ref{fig:4EM3}\subref{fig:4tempEM3}. Its evolution reveals low contribution to the flow field over a long period but with distinct exceptions. With regard to the corresponding spatial eigenmode $\Phi_3$, see Figure \ref{fig:4POD}\subref{fig:4spatEM3}, high amplitudes of $a_{\rm 3}$ are associated with pronounced formation of a large vortical structure. As shown in Figure \ref{fig:4EM3}\subref{fig:4SSmaxEM3-b}, which corresponds to the marked event in Figure \ref{fig:4EM3}\subref{fig:4tempEM3}, the formation of a large recirculation vortex is observed. The high amplitude of $a_{\rm 3}$ corresponds to a high-lift event as also discussed concerning Figure \ref{fig:3topoCCV_III}. In this way, the third eigenmode can be used as an identification method for extraordinary flow topologies.

\section{Discussion}\label{sec:5disc}
The variety of results presented in Sections \ref{sec:1inflow}--\ref{sec:4pod} is eventually used to draw a unified picture of observations, interactions and dependencies involved in the formation of light dynamic stall about an airfoil exposed to a periodic, three-dimensional gust. Following typical flow properties in wind energy, the free stream features a turbulence intensity of $Ti \approx 3~\%$ that manifests as small scale vortices. Though vortices increase the receptivity for perturbations in general, in Section \ref{sec:2topo} it is assessed to have a minor effect on the general flow formation. However, the following conclusions are to be seen in the light of a turbulent inflow that might obscure some connections.

The main question raised by the obtained results is: how does a three-dimensional, periodic inflow induce a counterclockwise dynamic polar along with a lift overshoot of $16~\%$. A linear connection between AoA and lift is ruled out in Section \ref{sec:3aero} because superposition of AoA still yields a counterclockwise dynamic polar (cf. Figure \ref{fig:2alphaCl}). Other reported flow situations resulting in a counterclockwise dynamic polar show only slight lead of AoA towards lift under experimental conditions characterized by high reduced frequency or low AoA \cite{Piziali1994}. In cases where AoA significantly leads, often additional phenomena take effect, such as flutter \cite{Rival2010} or a combination of pitch and plunge \cite{Cebeci2005,Leishman2016}. Since such phenomena are not subject of the present study, we hypothesize that a non-local coupling of the flow, particularly within the recirculation region, causes inhibited formation of stall and yields higher lift generation than expected. This hypothesis is reflected by measured aerodynamics as subsequently elucidated.

The surface bound flow upstream of separation interacts in spanwise direction via the pressure field. According to Figure \ref{fig:3phiClphase}, local AoI at $z/c=-0.18$ peaks roughly $\Delta t/T\approx0.26$ prior to local $c_{\rm L,loc}$ at the centerline. Once stall occurs for $t/T>0.4$ at $z/c=-0.18$, local AoA is still small at $z/c=0$ (see also Figure \ref{fig:1InflowAlphaV}). In turn, measured local airfoil bound flow is stabilized by the region of smaller instantaneous AoA enabling total lift further following the increase of local AoA. The decrease in total lift is accordingly inhibited after the turning point of local AoA.

Stage III of the dynamic stall cycle is characterized by maximum extent of separation region which in general comes along with a drop in lift. Instead, maximum total lift generation is obtained during AoA downstroke of local inflow (cf. Figures \ref{fig:2topo}\subref{fig:2topoMax-c} and \ref{fig:3phiCl}). Significant spanwise motion of flow is observed within the recirculation region during stall, as presented in Figures \ref{fig:3topoZ}\subref{fig:3topoMaxZ-c}. This enables a direct advective coupling of flow over a spanwise range on the airfoil. A manifestation of this type of coupling is the formation of vortices that is promoted in the stalled region (cf. Figure \ref{fig:3topoCCV_III}). Small, local and single vortices cannot significantly affect global lift. The emergence of single counterclockwise rotating vortices close to the trailing edge reflecting in total lift (cf. Figure \ref{fig:3topoCCV_III}) thus indicates the presence of a large-scale flow structure extending spanwise.

An analysis of large flow structures based on POD of flow topology supports the stated hypothesis. Although POD is derived from local flow formation, the temporal contribution of its most energetic spatial eigenmode follows global lift evolution even in details. Since the first eigenmode spatially represents the formation of separated flow, this indicates that the stall region interacts along the span of the airfoil which is in accordance with the argument of potentially strong spanwise coupling during stall.

\section{Summary}\label{sec:6concl}
The present study aims at experimentally investigating the impact of a three-dimensional, periodic gust with streamwise and spanwise variation on a DU\hbox{~}91\hbox{-}W2\hbox{-}250 wind profile undergoing light dynamic stall in a wind tunnel. Complementary experimental methods are used ranging from local flow visualization employing TRSPIV, through local lift measurements by means of surface pressure distribution, to global lift measurements obtained by a multi-component scale. Simultaneous application of all methods enables insight into involved fluid mechanic principles.

The three-dimensional, periodic gust leads to formation of light dynamic stall that is characterized by a counterclockwise dynamic polar and, during angular downstroke, lift augmentation by $16~\%$ compared to maximum static lift. These observations are traced back to (i) a stabilization of the surface bound flow by different AoA simultaneously acting on the airfoil that enable higher effective AoA and (ii) a non-local coupling of the stalled region in spanwise direction that inhibits stall. Both effects contribute to an increase of circulation around the airfoil reflecting lift.

As a next step, this insight into the formation of three-dimensional light dynamic stall needs to be verified by testing other reduced frequencies. Emerging stall cells and larger aerodynamic lag coming along with increased reduced frequency have to be considered in such investigation.

In the present study, a POD of the velocity fields above the airfoil at one spanwise location shows to capture total lift as well as local extreme flow events. This implies that experimental effort can be reduced in aerodynamic situations where three-dimensional inflow is involved. A combination of POD and TRSPIV aerodynamic data yields additional global information based on a local measurement only. On the other hand, POD can be used as a selection tool in complex flow situations. Extraordinary flow events are easily identified by the POD development coefficients. Eventually, a POD can be used as a low order model implemented in efficient engineering tools that wind turbine developers use for the design of rotor blades.

\section{Literature}
\bibliography{AirfoilPerformance}

\end{document}